\documentclass{natureprintstyle}

\usepackage{graphicx}
\usepackage{epsfig}
\usepackage{amssymb}
\usepackage{amsmath}
\usepackage{mathrsfs}
\usepackage{color}
\usepackage{url}

\newcounter{firstbib}

\newcommand\figcaption{\def\@captype{figure}\caption}

\newcommand{\ergs}{${\rm erg \ cm^{-2} \ s^{-1}}$ }

\newcommand{\todo}{\ifmmode {\Huge \bullet} \else {\Huge$\bullet$}\fi}

\newcommand{\kms}{\ifmmode {\rm km\,s}^{-1} \else km\,s$^{-1}$ \fi}
\newcommand{\cc}{\hbox{cm$^{-3}$}}

\newcommand{\cmii}{\hbox{cm$^{-2}$}}
\newcommand{\ergcms}{\ifmmode {\rm ergs\,cm}^{-2}\,{\rm s}^{-1} \else ergs\,cm$^{-2}$\,s$^{-1}$\fi}
\newcommand{\ergcmsA}{\ifmmode{\rm ergs}\, {\rm cm}^{-2}\,{\rm s}^{-1}\,{\rm\AA}^{-1} \else ergs\, cm$^{-2}$\, s$^{-1}$\, \AA$^{-1}$\fi}
\newcommand{\ergcmsHz}{\ifmmode{\rm ergs\,cm}^{-2}\,{\rm s}^{-1}\,{\rm Hz}^{-1} \else ergs\,cm$^{-2}$\,s$^{-1}$\,Hz$^{-1}$\fi}
\newcommand{\phcms}{\ifmmode {\rm ph\,cm}^{-2}\,{\rm s}^{-1} \else ,ph\,cm$^{-2}$\,s$^{-1}$\fi}
\newcommand{\phcmsA}{\ifmmode {\rm ph\,cm}^{-2}\,{\rm s}^{-1}\,{\rm\AA}^{-1} \else ph\,cm$^{-2}$\,s$^{-1}$\,\AA$^{-1}$\fi}

\newcommand\Msun{\ifmmode M_{\odot} \else $M_{\odot}$\fi}
\newcommand\msun{\ifmmode M_{\odot} \else $M_{\odot}$\fi}
\newcommand\Lsun{\ifmmode L_{\odot} \else $L_{\odot}$\fi}
\newcommand\Zsun{\ifmmode Z_{\odot} \else $Z_{\odot}$\fi}
\newcommand\mpyr{\ifmmode \Msun\,{\rm yr}^{-1} \else $\Msun\,{\rm yr}^{-1}$ \fi}

\newcommand{\Luv}{\ifmmode L_{1450} \else $L_{1450}$\fi}
\newcommand{\Lop}{\ifmmode L_{5100} \else $L_{5100}$\fi}
\newcommand{\Lthree}{\ifmmode L_{3000} \else $L_{3000}$\fi}
\newcommand{\lledd}{\ifmmode L/L_{\rm Edd} \else $L/L_{\rm Edd}$\fi}
\newcommand{\ledd}{\ifmmode L_{\rm Edd} \else $L_{\rm Edd}$\fi}
\newcommand{\lamLlam}{\ifmmode \lambda L_{\lambda} \else $\lambda L_{\lambda}$\fi}
\newcommand{\lbol} {\ifmmode L_{\rm bol} \else $L_{\rm bol}$\fi}
\newcommand{\llbol}{\ifmmode \log\left(\lbol/\ergs\right) \else $\log\left(\lbol/\ergs\right)$\fi}

\newcommand{\fuv}{\ifmmode f_{\lambda}\left(1450\AA\right) \else $f_{\lambda}\left(1450 {\rm \AA}\right)$\fi}
\newcommand{\fthree}{\ifmmode f_{\lambda}\left(3000\AA\right) \else $f_{\lambda}\left(3000{\rm \AA}\right)$\fi}
\newcommand{\fH}{\ifmmode f_{\lambda}\left(1.65\micron\right) \else
$f_{\lambda}\left(1.65\micron\right)$\fi}

\newcommand{\mbh}{\ifmmode M_{\rm BH} \else $M_{\rm BH}$\fi}
\newcommand{\lmbh}{\ifmmode \log\left(\mbh/\Msun\right) \else $\log\left(\mbh/\Msun\right)$\fi}

\newcommand \Hbeta {\ifmmode {\rm H}\beta \else H$\beta$\fi}
\newcommand \hb    {\ifmmode {\rm H}\beta \else H$\beta$\fi}
\newcommand  \mgii  {\ifmmode {\rm Mg}{\textsc{ii}} \else Mg\,{\sc ii}\fi}
\newcommand  \MGII  {\ifmmode {\rm Mg}\,{\sc ii}\,\lambda2798 \else Mg\,{\sc ii}\,$\lambda2798$\fi}
\newcommand  \siiv  {\ifmmode {\rm Si}\, {\sc iv}\ \else Si\,{\sc iv}\fi}
\newcommand  \SIIV  {\ifmmode {\rm Si}\,{\sc iv}\,\lambda1399 \else Si\,{\sc iv}\,$\lambda1399$\fi}
\newcommand  \civ  {\ifmmode {\rm C}\, {\sc IV}\ \else C\,{\sc IV}\fi}
\newcommand  \CIV  {\ifmmode {\rm C}\,{\sc iv}\,\lambda1549 \else C\,{\sc iv}\,$\lambda1549$\fi}
\newcommand  \NV  {\ifmmode {\rm N}\,{\sc v}\,\lambda1240 \else N\,{\sc v}\,$\lambda1240$\fi}
\newcommand  \nv  {\ifmmode {\rm N}\,{\sc v}\ \else N\,{\sc v}\fi}
\newcommand  \cv  {\ifmmode {\rm C}\,{\sc v}\ \else C\,{\sc v}\fi}
\newcommand  \LyA  {\ifmmode {\rm Ly}\,{\sc $\alpha$}\,\lambda1216 \else Ly\,{\sc $\alpha$}\,$\lambda1216$\fi}
\newcommand  \lya {\ifmmode {\rm Ly}\,{\sc $\alpha$}\ \else Ly\,{\sc $\alpha$}\fi}
\newcommand  \feii {\ifmmode {\rm Fe}\,{\textsc{ii}}\, \else Fe\,{\sc ii}\fi}

\newcommand  \aliii  {\ifmmode {\rm Al}{\textsc{iii}} \else Al\,{\sc iii}\fi}
\newcommand  \ciii     {C\,{\sc iii}}
\newcommand  \CIII  {\ifmmode {\rm C}\,{\sc iii]}\,\lambda1909 \else C\,{\sc iii]}\,$\lambda1909$\fi}
\newcommand  \oi    {\ifmmode \left[{\rm O}\,{\textsc i}\right] \else [O\,{\sc i}]\fi}
\newcommand  \OI    {\ifmmode \left[{\rm O}\,{\textsc i}\right]\,\lambda6300 \else [O\,{\sc i}]$\,\lambda6300$ \fi}

\newcommand  \oii   {\ifmmode \left[{\rm O}\,{\textsc ii}\right] \else [O\,{\sc ii}]\fi}
\newcommand  \OII   {\ifmmode \left[{\rm O}\,{\textsc ii}\right]\,\lambda3727 \else [O\,{\sc ii}]\,$\lambda3727$ \fi}
\newcommand  \oiii  {\ifmmode \left[{\rm O}\,{\textsc iii}\right] \else [O\,{\sc iii}]\fi}
\newcommand  \OIII  {\ifmmode \left[{\rm O}\,{\textsc iii}\right]\,\lambda5007 \else [O\,{\sc iii}]\,$\lambda5007$\fi}

\newcommand{\lmg}{\ifmmode L\left(\mgii\right) \else $L\left(\mgii\right)$\fi}
\newcommand{\fwmg}{\ifmmode {\rm FWHM}\left(\mgii\right) \else FWHM(\mgii)\fi}
\newcommand{\fwciv}{\ifmmode {\rm FWHM}\left(\civ\right) \else FWHM(\civ)\fi}
\newcommand{\fwhm}{\ifmmode {\rm FWHM} \else FWHM\fi}

\def\ergs{\rm erg~s^{-1}}

\def\civ{C {\sc iv}}
\def\feii{Fe {\sc ii}}
\def\mgii{Mg~{\sc ii}}
\def\oiii{[O~{\sc iii}]}

\begin{document}

\title{The properties of broad absorption line outflows based on a large sample of quasars}
\author{
Zhicheng He$^{1,2,3}$\thanks{E-mail: zcho@ustc.edu.cn},
Tinggui Wang$^{1,2}$\thanks{E-mail: twang@ustc.edu.cn}, 
Guilin Liu$^{1,2}$\thanks{E-mail: glliu@ustc.edu.cn}, 
Huiyuan Wang$^{1,2}$,
Weihao Bian$^{4}$,
Kirill Tchernyshyov$^{3}$,
Guobin Mou$^{1,2,5}$,
Youhua Xu$^{6}$,
Hongyan Zhou$^{1,2}$,
Richard Green$^{7}$
and Jun Xu$^{1,2}$
}
\maketitle

\begin{affiliations}
\item{CAS Key Laboratory for Research in Galaxies and Cosmology, Department of Astronomy, University of Science and Technology of China, Hefei, Anhui 230026, China}

\item{School of Astronomy and Space Science, University of Science and Technology of China, Hefei 230026, China}  

\item{Department of Physics \& Astronomy, Johns Hopkins University, Bloomberg Center, 
3400 N. Charles St., Baltimore, MD 21218, USA}
 
\item{Department of Physics and Institute of Theoretical Physics, Nanjing Normal University, Nanjing 210023, China}

\item{School of Physics and Technology, Wuhan University, Wuhan 430072, China}

\item{CAS Key Laboratory of Space Astronomy and Technology, National Astronomical Observatories, Beijing 100012, China}

\item{Steward Observatory, University of Arizona, Tucson, AZ, 85721-0065, USA}

\end{affiliations}

\begin{abstract}
Quasar outflows carry mass, momentum and energy into the surrounding environment, and 
have long been considered a potential key factor in regulating the growth of supermassive black holes
and the evolution of their host galaxies\cite{scannapieco2004, Murray2005,
ciotti2009, ostriker2010}.
A crucial parameter for understanding the origin of these outflows and measuring their influence on 
their host galaxies is the distance ($R$) between the outflow gas and the galaxy center
\cite{hopkins2009,di2005}. 
While $R$ has been measured in a number of individual galaxies
\cite{arav2008,moe2009,hamann2011,borguet2012a,arav2015,chamberlain2015,Arav2018,Xu2018,he2017a}, 
its distribution remains unknown.
Here we report the distributions of $R$ and the kinetic luminosities of quasars outflows, 
using the statistical properties of broad absorption line variability in a sample of 915 quasars from 
the Sloan Digital Sky Surveys.
The mean and standard deviation of the distribution of $R$ are $10^{1.4\pm0.5}$~parsecs.
The typical outflow distance in this sample is tens of parsec, which is beyond the theoretically predicted location
($0.01\sim 0.1$ parsecs) where the accretion disc line-driven wind is launched~\cite{murray1995,proga2000},
but is smaller than the scales of most outflows that are derived using the
excited state absorption lines~\cite{arav2008,moe2009,hamann2011,borguet2012a,arav2015,chamberlain2015,Arav2018,Xu2018}.
The typical value of the mass-flow rate is of tens to a hundred solar masses per year, or
several times the accretion rate.
The typical kinetic-to-bolometric luminosity ratio is a few per cent,
indicating that outflows are energetic enough to influence the evolution of their host galaxies.

\end{abstract}

Nowadays, theoretical models for galaxy formation and evolution routinely invoke the concept of "quasar feedback"--
the strong effect that the active supermassive black hole's (SMBH) energy output exerts on its host galaxy--
to keep massive galaxies from forming many stars and becoming overly massive.
In 10-40\% of the quasars in which the central source and outflowing gas are both in the line of sight, outflows may manifest 
themselves as broad absorption lines (BALs)\cite{gibson2009,allen2010}
and BAL outflows are therefore a candidate agent of quasar feedback.
The importance of outflows to active galactic nucleus (AGN) feedback can be quantified using 
the mass-flow rate ($\dot{M}_{\rm out}$) and the kinetic luminosity ($\dot{E_{\rm k}}$) of the outflowing material.
The $\dot{M}_{\rm out}$ and the $\dot{E_{\rm k}}$ of a BAL outflow can be
estimated from the distance ($R$) between the outflowing gas and the galaxy center, the total hydrogen column density
$N_{\rm H}$ and the fraction $\Omega$ of the solid angle subtended by the outflowing gas. 
Because the ionization parameter $U_{\rm H}$ of the plasma is inversely
proportional to the product of hydrogen number density $n_{\rm H}$ and $R^2$, i.e., $U_{\rm H}\propto 1/(n_{\rm H}R^{2}$),
$R$ can be obtained by measuring $U_{\rm H}$ and $n_{\rm H}$.
In general, $n_{\rm H}$ can be determined from the absorption lines of the excited states 
of ions (e.g., Fe II*, Si II*, S IV*), but this method is hindered by line blending and is therefore only
applicable to quasars with relatively narrow absorption lines.
During the last decade or so, outflow distances have been measured for only about a dozen of individual quasars
using this method\cite{arav2008,moe2009,hamann2011,borguet2012a,arav2015,chamberlain2015,Arav2018,Xu2018},
while the distributions of primary properties of BAL outflows remain in unknown.
In this work, we present a novel method to determine $R$ by examining the variability of BAL troughs.

BAL troughs vary on timescales ranging from several days to years~\cite{capellupo2011,capellupo2012,ak2012, ak2013,he2015,hemler2018}. 
There are two possibles causes of BAL variability, i.e. the tangential movement of the absorbing gas and changes in
the ionizing radiation incident on the gas. 
In the latter case, the variability timescales set a constraint on the
ionization or recombination timescales, which depend solely on the incident ionizing continuum
and gas density.
A series of statistical investigations on BAL variabilities have been
performed using the large multi-epoch spectroscopic dataset of Sloan Digital Sky Surveys (SDSS) DR10 and DR12 in
ref.\cite{wang2015,he2017}.
Ref.\cite{ak2013, wang2015} found that the majority of BAL variability is driven by variation of the ionizing continuum, 
and ref.\cite{he2017} further pinned down this fraction to be at least 80\% of BALs. 
On the basis of the above works, we focus on BAL variability driven by ionizing continuum variation and use it to derive 
the primary physical properties of BAL outflows.

The ionization state of a gaseous outflow demands for a period of time (the recombination timescale, $t_{rec}$~\cite{barlow1992,
krolik1995, wang2015}) to respond to changes in the ionizing continuum.
The gas ionization is connected to the average intensity of ionizing continuum over $t_{rec}$,
while the change in the average intensity of ionizing continuum between a pair of observations decreases
when the $t_{rec}$ increases (see Supplementary Figure 1 in Methods for details). 
In principle, an absorption line should only vary from one observation to another if the $t_{rec}$ is shorter than the timescale
on which the ionizing continuum varies and the time interval $\Delta T$ between the observations~\cite{barlow1992}.

The underlying distribution of the recombination timescale $t_{rec}$ (hereafter abbreviated as $t_{r}$) of the outflow 
gas, i.e., $f(t_{r})$, determines the fraction $F(\Delta T)$ of variable BAL troughs that
can be detected from a given large BAL quasar sample. 
We denote the probability of detecting the variability of a BAL with recombination timescale $t_{r}$ at $\Delta T$ as $p(t_{r}, \Delta T)$. 
Considering the ideal case, the BAL variability can (not) be detected when the recombination timescale $t_{r}$ is shorter (longer) 
than the observational time interval $\Delta T$.
In this case, $p(t_{r}, \Delta T)$ is a step function, i.e., $p=1$ for $t_{r}\le \Delta T$ and $p=0$ for $t_{r}> \Delta T$.
Hence, we can describe the ideal fraction curve as follows:
\begin{eqnarray} \label{eq1}
F_{ideal}(\Delta T) = \int_{0}^{+\infty} p(t_{r},\Delta T)f(t_{r})~d t_{r} = \int_{0}^{\Delta T}f(t_{r})~d t_{r}.
\end{eqnarray}
In reality, $p(t_{r}, \Delta T)$ is not a standard step function. 
and $F(\Delta T)$  depends not only on $\Delta T$ but also on the detection threshold. 
Given a certain detection threshold, 
we can write the actual fraction curve as the ideal fraction curve multiplied by a correction factor,
i.e., $F(\Delta T)=K(\Delta T)F_{ideal}(\Delta T)$. 
Note that because $p(t_{r}, \Delta T)$ is not a standard step function, 
the actual $F(\Delta T)$ may deviate from $F_{ideal}(\Delta T)$, even in the unrealistic case of complete 
detection of all variability.

Measuring $F(\Delta T)$ is sufficient for deriving the underlying distribution $f(t_{r})$. 
If $K(\Delta T)$ is constant, the underlying recombination
timescale distribution $f(t_{r})$ can be readily obtained by taking the derivative of $F(\Delta T)$ with
respect to $\Delta T$. If $K(\Delta T)$ is not constant, the derivative of $F(\Delta T)$
will deviate from $f(t_{r})$. 
However, in the Methods, we have performed a simulation showing that this deviation 
to be likely negligible for our sample.

$F(\Delta T)$ can be practically measured using the following method. Assuming that
we have a sample of $N$ multiply-observed BAL quasars which have already been sorted according to the rest time
interval between each pair of observations, we divide these $N$ quasars into $B$ bins, each of which contains approximately
the same number of objects. $\Delta T_i$ is the mean time interval
between the pair of observations for all quasars contained in the $i$-th bin. 
The fraction $F_i\equiv F(\Delta T_i)$ is measured to be
$F_i=k_i/N_i$,
where $k_i$ is the number of quasars with variable BAL, $N_i$ is the number of quasars contained in
the same bin, and we have dropped $\Delta T_i$ for simplicity. 
Assuming that the detections of BAL variability are mutually independent in the $i$-th bin, the probability of detecting
BAL variability follows a binomial distribution. The standard deviation of $k_i$ is
$\sigma_{k_i}=\sqrt{ k_i(N_i-k_i)}$. As a result, one can use $\sigma_{F_i}=\sqrt{ F_i(1-F_i)/N_i }$
as an estimate of the measurement error of $F_i$. Due to the incomplete independence of spectral pairs,
the actual uncertainty of the variable BAL fraction is larger than
that estimated from a binomial distribution.

The $F(\log_{10}\Delta T)$ curve measured from the SDSS sample (see Supplementary Figure~2) 
is presented in panel \textbf{a} of Figure~1. The fits of \civ~BAL troughs and the identification of \civ
BAL trough variabilities are shown in Methods and Supplementary Figure~3. 
The distributions of the parameters of the \civ~BAL troughs are shown in Supplementary Figure~4.
We assume that the logarithmic recombination time distribution is a Gaussian function G($t_c,t_\sigma$),
where $t_c$ and $t_\sigma$ are the mean and standard deviation of the Gaussian distribution, respectively.
Thus, the cumulative distribution function (CDF) of the Gaussian distribution is used to model the
fraction curve,
\begin{eqnarray}  \label{eq2}
F(t)= p_{0}\left[1+ {\rm erf}\left(\frac{t-p_{1}}{\sqrt{2}p_{2}}\right)\right],
\end{eqnarray}
where $t\equiv \log_{10}\Delta T$ is the logarithmic time interval between each pair of observations,
${\rm erf}(t)=1/\sqrt \pi \int_{-t}^{t}e^{-x^2} dx$ is the error function, and~$p_1=t_{c}$, $p_2=t_{\sigma}$, i.e.,
the mean and standard deviation of the Gaussian distribution. 
The reduced $\chi^2$ of the best-fit model is 1.18.
The best-fit mean and standard deviation of the Gaussian distribution are $t_{c}=0.36\pm0.14$ and $t_{\sigma}=1.01\pm0.22$.
The recombination timescale distribution is shown in panel \textbf{b} of Figure~1.
Taking the above recombination timescale distribution as the input Gaussian, and employing the photoionization model
(Supplementary Figure 5), we conduct a simulation test (see Methods) and generate the recovered Gaussians. 
The input and recovered Gaussians (Supplementary Figure 6) are consistent within $1\sigma$ uncertainty.

Ref.\cite{wang2015} found that \civ, \siiv, and \nv~respond negatively to an increasing
ionization parameter and then constrained the ionization parameter ${\log_{10}}~U$ of most BAL outflows to
be greater than 0 using photoionization simulations (see Figure 11 in ref.\cite{wang2015}). In view of this, we assume
${\log_{10}}~U = 0$ for all the objects in our sample and perform our subsequent calculations accordingly,
though we also report results based on other values of the ionization parameter for reference.

The recombination timescale $t_{r}$ of the \civ~line is related to the electron density $n_{e}$ and the
recombination rate $\alpha$ (see Methods for details).
According to the measured $t_{r}$ distribution, the mean and standard deviation of the
electron density distribution are $n_{e}=10^{6.19\pm1.02}$,~$10^{6.79\pm1.02}$,~$10^{4.79\pm1.02} \rm~and~10^{3.79\pm1.02} \cc$ at
$\log_{10}U=-2, -1 ,0 \rm~and~1$, respectively. 

The outflow distance $R$ can be determined as long as the ionization state and density are known (see Methods for details). 
As shown in Figure~2, the mean and standard deviation
of the $R$ distributions are $10^{1.70\pm0.54}, 10^{0.91\pm0.54}, 10^{1.41\pm0.54} \rm~and~10^{1.41\pm0.54}$ pc
at $\log_{10}U=-2, -1 ,0 \rm~and~1$, respectively. Our result that
the typical outflow radius is tens of pc indicates that
the BAL outflow locations is outside the theoretically predicted trough forming region
($0.01\sim 0.1 {\rm pc}$) for accretion disc line-driven winds~\cite{murray1995,
proga2000}, but are smaller than the scales of most outflows that are derived using the
excited state absorption lines~\cite{arav2008,moe2009,hamann2011,borguet2012a,arav2015,chamberlain2015,Arav2018,Xu2018}.

The mass-flow rate $\dot{M}_{\rm out}$ and kinetic luminosity $\dot{E_{\rm k}}$ are the key parameters to
quantify the powerfulness of the feedback effect. 
As shown in Figure~2, the mean and standard deviation of the distribution of $\dot{M}_{\rm out}$
(see Methods and Supplementary Figure 7, 8 for details)
are $10^{-0.09\pm0.70}, 10^{-0.27\pm0.70}, 10^{1.61\pm0.70}, 10^{2.86\pm0.70}$ $\mpyr$ at $\log_{10}U=-2, -1 ,0 \rm~and~1$, respectively.
At $\log_{10}U=0$, the typical value of $\dot{M}_{\rm out}$ is of ten to one hundred $\mpyr$. 

The ratio of the mass-flow rate to accretion rate $\dot{M}_{\rm out}/\dot{M}_{\rm acc}$ is a proxy
for exploring the relationship between the accretion system on small scales and the BAL outflows
on relatively large scales. The accretion rate is given by $\dot{M}_{\rm acc}=\lbol/\eta c^2$, where
$\eta = 0.1$ is the energy conversion efficiency. 
The mean and standard deviation of the distribution of $\dot{M}_{\rm out}/\dot{M}_{\rm acc}$
are $10^{-0.84\pm0.70}, 10^{-1.02\pm0.70}, 10^{0.86\pm0.70} \rm and~10^{2.10\pm0.70}$ at $\log_{10}U=-2, -1 ,0 \rm~and~1$, respectively.
At $\log_{10}U=0$, the typical $\dot{M}_{\rm out}$ is a few times of $\dot{M}_{\rm acc}$.
It is worth mentioning that ref.\cite{nomura2016} performed a series of two-dimensional
radiation-hydrodynamical simulations of line-driven disc winds for black holes with masses in the range
$\mbh =10^{6-9}\msun$ and Eddington ratios in the range $\varepsilon=0.1-0.5$. Their simulations predict
that $\dot{M}_{\rm out}$ can become comparable to $\dot{M}_{\rm acc}$
when the Eddington ratio $\varepsilon$ is at least 0.3.

As shown in Figure~2, the mean and standard deviation of the kinetic-to-bolometric luminosity ratio $\dot{E_{\rm k}}/\lbol$
(see Methods and Supplementary Figure 7, 8 for details)
are $10^{-2.82\pm0.70},10^{-3.01\pm0.70},10^{-1.11\pm0.70} \rm and~10^{0.15\pm0.70}$ 
at $\log_{10}U=-2, -1 ,0 \rm~and~1$, respectively.
At $\log_{10}U=0$, the typical value of $\dot{E_{\rm k}}/\lbol$ is a few percents. 
The threshold of $\dot{E_{\rm k}}/\lbol$ for effective AGN feedback is still under debate.
According to ref\cite{di2005}, a quasar outflow can effectively suppress
star formation in the host galaxy by directly expelling the ISM when $\dot{E_{\rm k}}/\lbol$
is at $10^{-1.3}$, i.e, 5\%. Meanwhile, ref.\cite{hopkins2009} proposed a “two-stage” feedback model.
In their model, dense clouds expand when the outflow passes by, which increases the clouds' cross-section and 
makes them more susceptible to radiative momentum driving and ionization heating by the quasar. 
For this case, an $\dot{E_{\rm k}}/\lbol$ value of $10^{-2.3}$ (i.e., $0.5\%$) is able to produce
enough feedback to suppress star formation in the host galaxy.
In either case, a large fraction of the BAL outflows in our study appear powerful enough to regulate the 
growth of the SMBHs and their host galaxies.

\begin{addendum}
 \item[Correspondence and request for materials]should be addressed to
   Z.-C. H or T.-G. W or G.-L. L (e-mail: zcho@ustc.edu.cn, twang@ustc.edu.cn, glliu@ustc.edu.cn).
\item[Acknowledgements] We acknowledge the financial support by the Strategic Priority Research Program "The Emergence of Cosmological 
Structures" of the Chinese Academy of Sciences (XDB09000000), NSFC (NSFC-11233002, NSFC-11421303, U1431229),
National Basic Research Program of China (grant No. 2015CB857005), National Science Foundation of China
(nos. 11373024, 11233003 \& 11873032) and National Key Research and Development Program of China (No. 2017YFA0402703).

Z.-C. H. is supported by China Scholarship Council (CSC, NO. 201706340030) during his stay at the Johns Hopkins University.
G.-L. L. is supported by the National Thousand Young Talents Program of China, and acknowledges the grant from the National Natural Science Foundation of China (No. 11673020 and No. 11421303) and the Ministry of Science and Technology of China (National Key Program for Science and Technology Research and Development, No. 2016YFA0400700).
G.B.M was supported by the National Natural Science Foundation of China (No. 11703022), and the Fundamental Research Funds for the Central Universities (WK2030220017). 

Funding for the Sloan Digital Sky Survey IV has been provided by the Alfred P. Sloan Foundation, the U.S. Department of Energy Office of Science, and the Participating Institutions. SDSS-IV acknowledges
support and resources from the Center for High-Performance Computing at
the University of Utah. The SDSS web site is www.sdss.org.

SDSS-IV is managed by the Astrophysical Research Consortium for the 
Participating Institutions of the SDSS Collaboration including the 
Brazilian Participation Group, the Carnegie Institution for Science, 
Carnegie Mellon University, the Chilean Participation Group, the French Participation Group, Harvard-Smithsonian Center for Astrophysics, 
Instituto de Astrof\'isica de Canarias, The Johns Hopkins University, 
Kavli Institute for the Physics and Mathematics of the Universe (IPMU) / 
University of Tokyo, Lawrence Berkeley National Laboratory, 
Leibniz Institut f\"ur Astrophysik Potsdam (AIP),  
Max-Planck-Institut f\"ur Astronomie (MPIA Heidelberg), 
Max-Planck-Institut f\"ur Astrophysik (MPA Garching), 
Max-Planck-Institut f\"ur Extraterrestrische Physik (MPE), 
National Astronomical Observatories of China, New Mexico State University, 
New York University, University of Notre Dame, 
Observat\'ario Nacional / MCTI, The Ohio State University, 
Pennsylvania State University, Shanghai Astronomical Observatory, 
United Kingdom Participation Group,
Universidad Nacional Aut\'onoma de M\'exico, University of Arizona, 
University of Colorado Boulder, University of Oxford, University of Portsmouth, 
University of Utah, University of Virginia, University of Washington, University of Wisconsin, 
Vanderbilt University, and Yale University.
 \item[Author contributions] Z.-C. H. presented the idea, made the calculations and wrote the manuscript.
T.-G. W., G. -L. L., H.-Y. W., W. H. B., G.-B. M., H.-Y. Z. and R. G. discussed the idea and the calculations.
Y.-H. X., K. T., T.-G. W., G. -L. L. and J. X revised the manuscript.
All authors discussed and gave comments on the contents of the paper.

 \item[Competing Interests] The authors declare that they have no
competing financial interests.
\end{addendum}

\begin{figure}[h]
\centering
\includegraphics[height=8.cm]{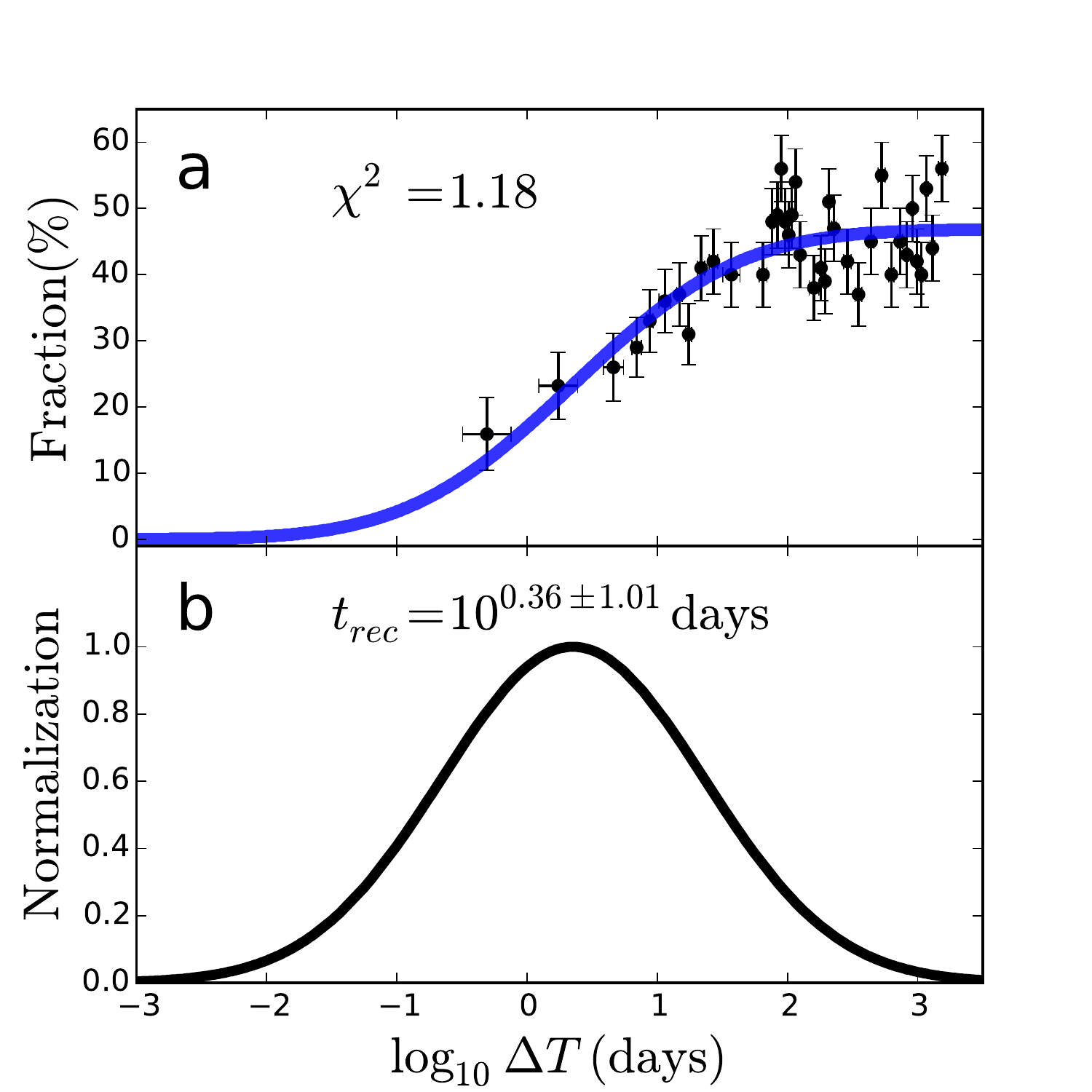}
\caption{\textbf{The fraction curve of BAL variabilities and the inferred recombination timescales
distribution in the SDSS sample.}
 \textbf{a}, the cumulative distribution function of a Gaussian distribution
is used to model the fraction curve of BAL variabilities. The vertical error bars mark the $1\sigma$ uncertainty of the
fraction curve and the horizontal one marks the width of each time bin.
\textbf{b}, the inferred Gaussian distribution of the recombination timescales of BAL outflow gas.
The mean and standard deviation of the recombination timescales are
$t_{rec}=10^{0.36\pm1.01}$ days. }
\label{fig:timebin}
\end{figure}

\begin{figure}
\centering
\includegraphics[height=11.5cm]{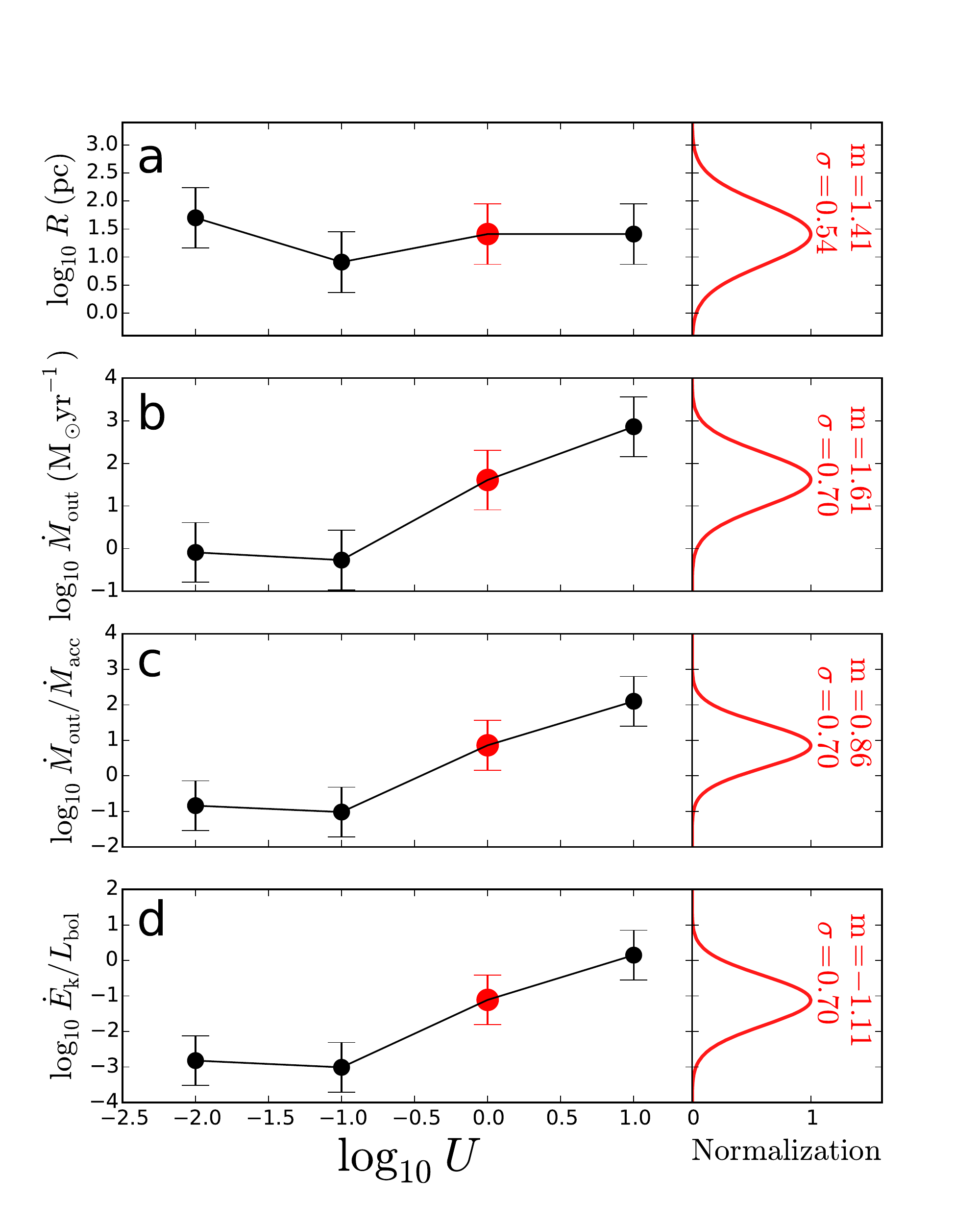}
\caption{\textbf{Distributions of the properties of the BAL outflows at different ionizing parameters.} 
The distributions at the ionizing parameter $\log_{10} U = 0$ (red points) are the final results 
(marked with the mean and standard deviation values in the right-hand panels). The error bars mark the
standard deviation of the distributions.
\textbf{a}, the distance distribution is $R$ = $10^{1.41\pm0.54}$ pc.
\textbf{b}, the mass-flow rate distribution is $\dot{M}_{\rm out}$ = $10^{1.61\pm0.70}$~$\rm \mpyr$. 
\textbf{c}, the mass-flow rate to accretion rate distribution is $\dot{M}_{\rm out}/\dot{M}_{\rm acc}$ = $10^{0.86\pm0.70}$.
\textbf{d}, the ratio of the kinetic-to-bolometric luminosity ratio distribution is $\dot E_{\rm k}$/$\lbol$ = $10^{-1.11\pm0.70}$.}
\label{fig:result}
\end{figure}

\begin{methods}

\subsection*{\textbf{The change in the average intensity of ionizing continuum.}}
For a typical high luminosity quasar, we assume that $\mbh=10^9\msun$
, $M_{i}=-26$, the relaxation timescale $\tau = 200$ days, and the value of the structure function
at infinity $\rm SF_{\infty}=0.2$ mag~\cite{macleod2010}. 
As shown in Supplementary Figure 1, we generate a light curve at 1500\AA~
from the damped random walk (DRW) model\cite{kelly2009, kozlowski2009, guo2017},
using the Python package astroML\cite{kelly2009}. 
The time interval between a pair of observations is assumed to be 30 days.
The change in the average intensity of ionizing continuum (averaged over a recombination
timescale $t_{r}$ before the observation) over the 100 days is obviously smaller than that of 10 days.
In general, the change in the average intensity of ionizing continuum between a pair of observations decreases
when the $t_{r}$ of gas is increased (panel \textbf{b} of Supplementary Figure 1).
In principle, the change in the average intensity of ionizing continuum can be ignored when
the $t_{r}$ is longer than the time interval between a pair of observations.

\subsection*{\textbf{BAL Quasar Sample.}}
We merge the BAL quasar catalog of SDSS data release 7 (DR7)\cite{shen2011}
and that of DR12\cite{paris2017}. 
Then we compare this catalog with SDSS data release 14 (DR14) and select quasars with multiple spectroscopic observations. 
To investigate the variability of the \CIV~ BAL trough, we adopt a redshift cut ($1.9 < z < 4.7$). 
To ensure detection of major absorption lines, we only keep quasars with at least one spectrum with signal to noise (S/N) at
SDSS $g$ band greater than 10. 
After these cuts, we obtain a sample of 1728 BAL quasars with spectra taken at two or more epochs. 
For each quasar with $m$ ($m \ge 2$) spectra, there are $C^2_m = m(m-1)/2$ spectral pairs. 
Then, there are total 9772 spectral pairs of 1728 BAL quasars in our sample. 
The amplitude of continuum variation is defined as
\begin{eqnarray}\label{eq3}
\frac{\Delta L}{L}=2\frac{L_2-L_1}{L_1+L_2},
\end{eqnarray}
where $L_1$ and $L_2$ are the fluxes measured at the first and second observations, respectively.
The continuum (1500\AA) variation amplitudes $|\Delta L/L|$ versus the rest time interval $\log_{10}\Delta T$ are plotted in gray
points in panel \textbf{a} of Supplementary Figure 2. The distribution of $\log_{10}\Delta T$ are shown in panel \textbf{b} of 
Supplementary Figure 2.
The distribution of $|\Delta L/L|$ are shown in panel \textbf{c} of Supplementary Figure 2. 
The fiber positions of BOSS quasar targets were purposefully offset in order to optimize the throughput of light at 4000\AA, 
while the standard stars used for flux calibration are positioned for 5400\AA. 
This results in a large uncertainty in the flux calibration of quasar spectra in the BOSS survey \cite{dawson2013}.
In DR14, Ref.\cite{abolfathi2018} have re-reduced BOSS spectra and improved the flux calibration by 
adding new atmospheric distortion corrections at the per-exposure level \cite{margala2016}.
We cut the $|\Delta L/L|$ at 10\% which is greater than the amplitude of spectrophotometric uncertainties of 6\% level\cite{shen2014}.
The $|\Delta L/L|$ distributions of our sample and the DRW model\cite{kelly2009, kozlowski2009, guo2017} 
of a typical quasar ($\mbh=10^9\msun$, $M_{i}=-26$) are consistent at $|\Delta L/L| \ge$ 10\% 
(Kolmogorov-Smirnov test: $r=0.02,~p=0.23$).
These two distributions with $|\Delta L/L|<$ 10\% are significantly different ($r=0.08,~p=1.6\times 10^{-14}$).

As shown in black points of panel \textbf{a} of Supplementary Figure 2, to keep the $|\Delta L/L|$ flat in different 
time bins, we only select those spectral pairs of $10\%< |\Delta L/L| <30\%$.  Among this sample, 
there are 3686 spectra pairs from 915 quasars (black points). Through the identification of variable absorption lines, 
1572 pairs (red points) of spectra from 432 BAL quasars are detected to have \civ~BAL variations.

As shown in panel \textbf{a} of Figure~1 and panel \textbf{d,~e,~f} of Supplementary Figure 2, 
in order to measure the fraction $F(\Delta T)$, the selected 3686 pairs are sorted 
according to the rest time interval between the observations and divided among 38 bins. Each bin contains
100 spectra pairs except the first three bins. The first bin has 44 spectral pairs of $\log_{10}\Delta T <0$.
The second bin has 68 spectral pairs of $0<\log_{10}\Delta T <0.5$. The third bin has 74 spectral pairs. 
We have marked the widths (standard deviation) of all the bins in panel \textbf{a} of Figure~1 and 
panel \textbf{d,~e,~f} of Supplementary Figure 2.

The S/N of spectra may affect the detectability of BAL variability. As shown in the right panel
of Supplementary Figure 2, the $S/N$ of the spectra at SDSS g band is nearly constant in all time intervals. 
The detectability of absorption line variability should therefore be approximately the same in the different time bins.  
In addition, there are two factors that may affect the
BAL variability: the amplitudes of continuum variations $|\Delta L/L|$ and the basic physical parameters
of the host quasars e.g., the bolometric luminosity $\lbol$ and the central SMBH mass $M_{\rm BH}$~\cite{he2015}.
The BAL variability increases with the amplitude of continuum variations while the timescale of continuum variations 
may increase with $\lbol$ and $M_{\rm BH}$. 
The amplitudes of continuum variations $|\Delta L/L|$ and the monochromatic luminosities at 1500~\AA~are
almost constant for all time bins. Thus, the influence of these two factors can also be ignored. 

\subsection*{\textbf{Fitting \civ~BAL trough.}}
To reliably characterize the continuum and delineate it from the \civ, \nv~BAL troughs, we use the unabsorbed quasar 
templates\cite{wang2015} drived from SDSS Data Release 7 (DR7) to fit the spectra. 
Following ref.\cite{wang2015,he2017}, we use a double power-law function (Equation 1 in ref.\cite{wang2015}) 
as the scale factor to scale these templates. 

Dividing the spectra by the continuum, we obtain the normalized spectrum and then mark the
contiguous deficient pixels as the possible intrinsic absorption lines region\cite{wang2015,he2017} of \civ~in 
the normalized spectrum. The marked region with a width of $\Delta \ln\lambda\geqslant 
10^{-3}$ (greater than 300 km~s$^{-1}$ in velocity) and statistically significant 
than 5$\sigma$ will be screened as the intrinsic moderate to broad absorption line.
Finally, we exclude the false ones (due to an improper fit in most cases) by the visual inspection.
The distributions of the weighted centroid velocity~\cite{ak2013} and 
width of \civ~BAL are shown in panel \textbf{a} and \textbf{b} of Supplementary Figure 4.
The equivalent width (EW) of the \civ~BAL troughs is calculated as follows:
$EW=\int[1-f_{obv}(\lambda)/f_{con}(\lambda)]d\lambda$. The integration is done for the identified
absorption line region.
The averaged EW for each object of the 915 quasars are shown in Supplementary Figure 5.

The normalized residual flux in the trough of a partially obscured absorber\cite{arav1999,hall2002} is
\begin{eqnarray}\label{eq4}
I(v)=1-C(v)+C(v)e^{-\tau(v)},
\end{eqnarray}
where $C(v)$ and $\tau(v)$ are the covering factor and the optical depth of the ion at velocity
$v$, respectively. 
The oscillator strengths of the blue and red components for the resonance doublet \civ~1548.2, 1550.8\AA~
are $f_{\rm blue}$ = 0.19 and $f_{\rm red}$ = 0.095, respectively. 
This renders an optical depth ratio 
$\tau_{\rm blue}/\tau_{\rm red}$=$(\lambda_{\rm blue} f_{\rm blue})/(\lambda_{\rm red} f_{\rm red})$
close to 2. Thus, we will use the doublet
components to fit the BAL troughs. The covering fraction $C$ has been found different at different 
velocities\cite{arav2013,leighly2018}. However, for simplicity, we only consider a constant covering factor
for the whole BAL trough and allow the optical depth $\tau$ to vary with velocity. Note that our result of the \civ~ 
column densities is a conservative estimation. According
to the partial covering model, we obtain a set of equations of $\tau(v_i)$ as follows:
\begin{eqnarray}\label{eq5}
\left\{  
\begin{array}{rcl}  
I_{v_{1}} & = & 1-C+Ce^{-\tau(v_{1})}, \\
 & \vdots &  \\
I_{v_{k}} &=&1-C+Ce^{-\tau(v_{k})}, \\
I_{v_{k+1}}&=&[1-C+Ce^{-\tau(v_{k+1})}][1-C+Ce^{-2\tau(v_{1})}], \\
 & \vdots &  \\
I_{v_{n-k}}&=&[1-C+Ce^{-\tau(v_{n-k})}][1-C+Ce^{-2\tau(v_{n-2k})}], \\
I_{v_{n-k+1}}&=&1-C+Ce^{-2\tau(v_{n-2k+1})}, \\
 & \vdots &  \\
I_{v_{n}}&=&1-C+Ce^{-2\tau(v_{n-k})},
\end{array}  
\right.  
\end{eqnarray}
where $\tau$ is the optical depth of red component. There are $n$ equations in total with $n-k+1$
unknown variables, where $k=5$ for a bin of 0.5\AA~in wavelength. Since there are more constraints
than unknown variables, the equation set has no exact solution. We therefore use the least-squares method
to find a set of $\{\tau(v_i)\}$ that best fits these equations. To account for the noise in the
flux and the uncertainty of continuum, the low-limit of the optical depth $\tau$ is set to $-0.1$
(corresponding to a normalized flux $I\simeq 1.1$). For most of the troughs, the best fit results give a
reduced $\chi^2$ around 1. An example fit is shown in panel \textbf{a} of Supplementary Figure 3.
Note that the optical depth for the case of saturated absorption must be underestimated.

After the troughs are fitted, the \civ~column densities are obtained by integrating the
optical depth over the troughs~\cite{savage1991}:
\begin{eqnarray}\label{eq6}
N_{ion} =\frac{3.7679\times10^{14}\cmii}{\lambda f}\int\tau(v)dv,
\end{eqnarray}
where $\lambda$ and $f$ are the transition's wavelength and oscillator strength, respectively,
and the velocity $v$ is measured in \kms. The \civ~column densities versus the BAL EW for 
the 915 quasars are shown in Supplementary Figure 5.

\subsection*{\textbf{Identification of the variable region of \civ~BAL trough.}}
As described in ref.\cite{wang2015,he2017}, in order to identify the variation region of \civ~BAL between 
a pair of spectra, we first need to exclude the influences of the continuum and emission line.
To account for the potential variations of the continuum shape, we select the higher S/N spectrum of the
pair of spectra as a template to match the other spectra by rescaling it using the
double power-law function (Equation 1 in ref.\cite{wang2015}).
To account for variations of the emission line, we add/subtract a Gaussian to/from the
rescaled spectrum. An example of the fit is displayed in panel \textbf{b} of Supplementary Figure 3.
Compared with the unabsorbed quasar template matching, the
rescaled template matching produces a better fit
outside the absorption line region in most cases. As a result, we will measure
the absorption line variability from the difference spectrum.

We take three steps to identify the variable absorption line components from 
the difference spectrum.
Firstly, we search for the contiguous negative and positive pixels and 
mark all pixels where the difference is greater than 3$\sigma$. 
we screen the adjacent marked pixels and connect them to form a variable region. 
Secondly, we expand such regions into neighboring pixels which have the same
sign but are less than 3$\sigma$ significant level. 
Finally, we merge the neighboring regions which have the same variable sign and 
a separation of less than four pixels. 
The confidence with which a region of BAL is assigned to be variable is defined as:
$N_{\sigma}=\rm \sum |\Delta flux|/ \sqrt{\sum \sigma^2}$,
where the flux uncertainties ($\rm \sigma= \sqrt{\sigma^2_{flux_{1}}+\sigma^2_{flux_{2}}}$)
of the two spectra includes the possible systematic uncertainties due to rescaling. 
We perform the identification of variable absorption in the wavelength coverage from 1410\AA~to 1500\AA~
(corresponding velocity $2.7\times 10^4 \kms$~to 0).
1572 pairs of spectra in 432 BAL quasars are detected (at $3\sigma$ detection threshold) 
to have \civ~BAL variations. 
As discussed in ref\cite{rogerson2018}, the coordinated variability in the troughs of different 
velocities are also found in our sample. 464 out of 1572 spectral pairs are detected two or more varied troughs. 
In the 464 spectral pairs, the varied troughs in 75\%(347/464) spectral pairs are coordinated.
This result also suggests that the BAL variabilities are likely due to
clouds at different velocities responding to the same changes in ionizing 
flux\cite{ak2012, ak2013, wang2015, he2017, rogerson2018}.
 
The distributions of the weighted centroid velocity and width 
of \civ~BAL for the 432 BAL quasars are plotted as orange line in panel \textbf{a} and \textbf{b} of Supplementary Figure 4.
There is no significant difference of the weighted centroid velocity distributions 
between the 915 quasars (black) and 432 quasars with varied BAL (Kolmogorov-Smirnov test: $r=0.06,~p=0.25$).
There is a weak difference of the BAL width distributions
between the 915 quasars (black) and 432 quasars with varied BAL ($r=0.09,~p=0.01$).
The distributions of varied BAL region widths for 1572 spectral pairs are shown in panel \textbf{c} of Supplementary Figure 4.
The number of occurrences of BAL absorption (black) and variable region (orange line) at different velocity bin are shown
in panel \textbf{d} of Supplementary Figure 4. The purple one is the percentage of BAL variability, 
i.e., the ratio of the orange one to the black one. 
The percentages of BAL variability increase with the velocities from 0 to $2\times 10^4 \kms$ which is similar
to the previous studies\cite{capellupo2011,capellupo2012,ak2013,rogerson2018}. 
The percentage of BAL variability is roughly constant
(even a slight drop) from $2\times 10^4 \kms$ to $2.7\times 10^4 \kms$ which is similar
to ref\cite{capellupo2011}. 
The variable regions are found across a wide range of velocities (mainly greater than $5.0\times 10^3 \kms$), 
suggesting that our results of the outflow distributions are the comprehensive statistical study for the
outflow with different velocities.

\subsection*{\textbf{Simulation test.}}
To validate our method, we measure $F(\Delta T)$ from a mock sample of $10^3$ quasar light curves
generated by the DRW model, then estimate $f(t_{r})$ and compare it with the input distribution of 
recombination timescales $t_{r}=10^{0.36\pm1.01}$ days.
Each light curve has $10^5$ evenly spaced points, with the time interval between two neighboring points 
corresponding to 0.1 day.

In the mock sample simulation, we use the 1500\AA~flux as an indicator of the ionizing flux, because
the highly concordant variations of the absorption lines and UV continuum strongly suggest
that the changes of the flux in the observed 1500\AA~rest-frame UV correlates with the changes of the
ionizing flux (see ref.\cite{wang2015,he2017}). 

We use the Cloudy~\cite{ferland2013} (version c13.03) to simulate the BAL EW variation responding to the 
variation of ionizing flux. Assuming that the distribution of optical depth $\rm \tau_{v}$ in the \civ~BAL trough is Gaussian
and the covering factor C=1, 
we calculate the BAL EW curve as a function of the column density $N_{\rm CIV}$ at different widths $\rm \sigma_{v}$ (color lines in 
panel \textbf{a} of Supplementary Figure 5). The measured BAL EW and $N_{\rm CIV}$ for the 915 quasars in our sample
are also plotted in panel \textbf{a} of Supplementary Figure 5.
The response of $N_{\rm CIV}$ and BAL EW to the variations of ionization parameters $\log_{10} U$ 
at the hydrogen column density $\log_{10}N_{\rm H}$ = 21 $\rm cm^{-2}$ and 22 $\rm cm^{-22}$ are shown in panel 
\textbf{b} of Supplementary Figure 5. According to the observed negative response of \civ~ BAL EW to the increasing ionization parameter,
we calculate the initial ionization parameter $\log_{10}U_1$ for the 915 quasars (black points in \textbf{b} of Supplementary Figure 5).
It shows that most of the initial ionization parameter is $\log_{10}U_1>-1$ at $\log_{10}N_{\rm H}$ = 21 $\rm cm^{-2}$,
and is $\log_{10}U_1>0$ at $\log_{10}N_{\rm H}$ = 22 $\rm cm^{-2}$. The BAL EW will respond to the $\log_{10} U$ along 
each color line at different widths $\rm \sigma_{v}$.
The mean fluxes over a recombination time scale $ {L_1}_{mean}$ and ${L_2}_{mean}$ determine the level of gas ionization state
at a pair of observations. 
According to the definition of the ionization parameter, we have ${L_2}_{mean}$/${L_1}_{mean}$ = $U_2/U_1$, 
i.e., $\log_{10}U_2$=$\log_{10}U_1$ + $\log_{10} ({L_2}_{mean}/{L_1}_{mean})$.
Given the $\log_{10}U_1$ and $\log_{10}U_2$, we can obtain the variation of BAL trough, i.e., $\Delta EW$.
Furthermore, we assume that variation of a BAL trough is detectable if $|\Delta EW|$ is greater than a certain threshold.
In our simulation, if the detection thresholds are $|\Delta EW|$ = 1.3\AA~(for $\log_{10}N_{\rm H}$ = 21 $\rm cm^{-2}$) and
$|\Delta EW|$ = 2.2\AA~(for $\log_{10}N_{\rm H}$ = 22 $\rm cm^{-2}$), the maximum value of the simulation fraction curves 
will be $\approx$ 50\% which is close to the observed fraction curve.

We compute $F(\log_{10}\Delta T)$ at values of $\log_{10} \Delta T$ that are spaced between $0.3$ and $3.9$ with a
stepsize of $0.1$. Note that the quasar variabilities significantly deviate from the DRW
model on the short timescale \cite{mushotzky2011,kasliwal2015}. And the DRW model can not generate enough large
amplitude of flux variability on $\log_{10} \Delta T <0.3$.
The detailed simulation processes are as follows:
\begin{itemize}

\item[1.] To produce one detection or non-detection of BAL variability, we generate a random recombination time $t_{r}$ 
from the input Gaussian distribution ($t_{c}=0.36\pm0.14$ and $t_{\sigma}=1.01\pm0.22$) 
and pick a random $\Delta L/L$ and BAL EW from the sample of 915 quasars.
We choose a random mock light curve and a pair of observation times separated by the current $\log_{10} \Delta T$,
average the light curve for a recombination time before each observation to obtain $ {L_1}_{mean}$ and ${L_2}_{mean}$.
Combined with the initial $\log_{10}U_1$ , $ {L_1}_{mean}$ and ${L_2}_{mean}$, we can obtain the $\log_{10} U_2$.
Then the variation of BAL trough $\Delta EW$ is generated.
If $|\Delta EW|$ is greater than the certain threshold (1.3\AA~for $\log_{10}N_{\rm H}$ = 21 $\rm cm^{-2}$,
2.2\AA~for $\log_{10}N_{\rm H}$ = 22 $\rm cm^{-2}$), then we will mark this realization as a variable.

\item[2.] Repeat step 1 for 1000 times per time interval, then measure the detection fraction at each
time interval.

\end{itemize}

Panel \textbf{a} of Supplementary Figure 6 shows the fractions $F(\log_{10}\Delta T)$ and the associated errors measured from
the mock sample. We use the CDF of a Gaussian distribution (see Eq.~\ref{eq2}) to 
model the fraction curves. The input and recovered Gausses are exhibited in panel \textbf{b} of Supplementary Figure 6.
The mean and standard deviation of the best fit parameters for $\log_{10}N_{\rm H}$ = 21 $\rm cm^{-2}$
are: $t_{c}=0.42\pm0.03$ and $t_{\sigma}=0.98\pm0.06$.
For $\log_{10}N_{\rm H}$ = 22 $\rm cm^{-2}$, the best fit parameters
are: $t_{c}=0.43\pm0.03$ and $t_{\sigma}=0.97\pm0.06$.
The parameters of the recovered Gausses are consistent with the parameters of the input Gauss within 1$\sigma$ uncertainty.
The agreement between the input and recovered Gaussian distributions indicates that our method is 
credible.

\subsection*{\textbf{The ionization parameter $U$.}}
The \civ~and \siiv~BAL troughs are both in our sample spectral coverage. 
However, it is difficult to measure the ionization parameter of the absorber when the \siiv~trough
disappears in the noise, or when the \civ~trough becomes too saturated to be measured. In addition,
the outflow may be multi-phase with a range of ionization states\cite{leighly2018}, 
as proposed by ref.\cite{arav2013} based on the analysis of far-UV absorption lines of the quasar HE 0238-1904, 
and warm absorbers of Seyfert galaxies~\cite{steenbrugge2009, detmers2011}.
In view of these issues, we only consider a fiducial ionization parameter for all the objects.
Ref.\cite{wang2015} found that \civ, \siiv, and \nv~respond negatively to an increasing
ionization parameter, implying an ionization parameter of ${\log_{10}}~U \ge 0$ (see Figure 13 in ref.\cite{wang2015})
for most BAL outflows. We therefore adopt the ionization parameter ${\log_{10}}~U = 0$ for all the objects.

\subsection*{\textbf{Hydrogen column density $N_{\rm H}$ under three AGN SED shapes.}}

To determine the hydrogen column density $N_{\rm H}$, we run a series of photoionization simulations
using Cloudy~\cite{ferland2013} (version c13.03). Since the gas ionization is insensitive
to the electron density at a given ionization parameter, we take a typical electron
density $n_{e}$ = $10^5~\cc$. 

The photoionization of an outflow depends on the incident SED.
In this work, we compare the photoionization solutions obtained using three different AGN SEDs:
MF87, UV-soft and HE 0238 (see Figure 10 of their paper\cite{arav2013}). 
The MF87 SED \cite{mathews1987}
is usually used to describe radio-loud quasars, whose most obvious feature is the so-called
big blue bump. The UV-soft SED is used for high-luminosity radio-quiet quasars~\cite{dunn2010}.
The HE 0238 SED is used to describe HE 0238-1904 ($z$ = 0.6309), which is a radio-quiet quasar~\cite{arav2013}.
In this work, the averaged value of the photoionization solutions for the three SEDs is adopted as our final result.

As mentioned in ref.\cite{arav2013}, AGN outflow gases have supersolar
metallicities (e.g., Mrk 279: $Z\simeq2\Zsun$\cite{arav2007}; SDSS J1512+1119:
$1\Zsun \leq Z \leq 4\Zsun$\cite{borguet2012b}; SDSS J1106+1939: Z=4\Zsun\cite{borguet2013}). 
We therefore adopt a moderate metallicity $Z = 2\Zsun$ in our calculations. 
Using the photoionization simulations, we obtain the relations between the hydrogen column density $N_{\rm H}$ and the
\civ~column density $N_{\rm CIV}$ at ${\log_{10}}~U = 0$ (see Supplementary Figure 7). 
Then, the hydrogen column density $N_{\rm H}$ of all the outflows can be estimated from these relations.

\subsection*{\textbf{The recombination timescale $t_{r}$ of \civ~line.}}
The recombination timescale $t_{r}$ \cite{arav2012} of the \civ~line is related to the electron density $n_{e}$ and the
recombination rate $\alpha$:
\begin{eqnarray}  \label{eq7}
t_{r}= \left[-\frac{\Delta L}{L} \alpha_{\rm \civ} n_{e}\left(\frac{n_{\rm \cv}}{n_{\rm \civ}}-\frac{\alpha_{\rm \ciii}}{\alpha_{\rm \civ}}\right)\right]^{-1}
\end{eqnarray}
, where $\Delta L/L$ is the amplitude of change in the incident ionizing flux.
Using the Chianti atomic database version 8.0~\cite{del2015} at a nominal temperature of $2\times 10^{4}$~K, we take 
the recombination rates $\alpha_{\rm \civ}=5.3\times 10^{-12}~\rm{cm^{3}~s^{-1}}$ (from \cv~to \civ) and  
$\alpha_{\rm \ciii}=2.1\times 10^{-11}~\rm{cm^{3}~s^{-1}}$ (from \civ~to \ciii).
At the ionization parameters $\log_{10}U=-2, -1 ,0 \rm~and~1$, the ratio of number densities of \cv~to~\civ~are 
$n_{\rm \cv}/n_{\rm \civ} \approx 0.3, 5, 100\rm~and~1000$, respectively.

\subsection*{\textbf{Calculation of the outflow properties.}}
The outflow distance $R$ can be determined as long as the ionization state and density are known. 
The ionization parameter is defined as follows:

\begin{eqnarray}\label{eq8}
U=\frac{Q_{\rm H}}{4\pi R^{2}n_{\rm H}c},
\end{eqnarray}
where $Q_{\rm H} = \int_{13.6~{\rm eV}}^{+\infty}\rm L_{E} dE$ is the source emission rate of
hydrogen-ionizing photons, $c$ is the speed of light, and $n_{\rm H} \approx 0.83n_{e}$ is the hydrogen number density.
Assuming that the distributions of $Q_{\rm H}$ and $n_{\rm H}$
are independent of each other, the distribution of $R=\sqrt{Q_{\rm H}/(4\pi U n_{\rm H}c)}$ is the product distribution of 
the above two distributions. 

If the outflow is in the form of a thin partial shell ($\Delta R/R\ll 1$), $\dot{M}_{\rm out}$ and 
$\dot{E_{\rm k}}$ can be given by~\cite{borguet2012a}
\begin{eqnarray}
\dot{M}_{\rm out} &=& 4\pi\Omega C R\mu m_{p} N_{\rm H}v,\label{eq9}\\
\dot{E_{\rm k}} &=& 2\pi\Omega C R\mu m_{p} N_{\rm H}v^3,
\label{eq10}
\end{eqnarray}
where $\mu$= 1.4 is the mean atomic mass per proton, $m_{p}$ is the mass of the proton, and $v$
is the radial velocity of the outflow. Here we adopt the weighted centroid velocity of the \civ~BAL
trough, i.e., the mean of the velocities where each data point is weighted by its distance from
the normalized continuum level~\cite{ak2013}. $\Omega$ and $C$ are the global covering factor and individual 
covering factor, respectively. Following the usual statistical approach
for \civ~BALs, the global covering factor is set to $\Omega$ $\simeq $ 0.2.

Combining Eq.~\ref{eq8} and Eq.~\ref{eq9}, the expression for $\dot{M}_{\rm out}$~is
\begin{eqnarray}\label{eq11}
\dot{M}_{\rm out} =4\pi\Omega C \mu m_{p} N_{\rm H}v \sqrt{\frac{Q_{\rm H}}{4\pi U n_{\rm H}c}}.
\end{eqnarray}
The $\dot{M}_{\rm out}$ distribution can be deduced from the distributions of $CQ_{\rm H}^{1/2}N_{\rm H}v$ 
(see Methods for details) and $n_{\rm H}$.

The accretion rate is $\dot{M}_{\rm acc}=\lbol/\eta c^2$, where
$\eta = 0.1$ is the energy conversion efficiency. From Eq.~\ref{eq11},
the $\dot{M}_{\rm out}/\dot{M}_{\rm acc}$ can be written as :
\begin{eqnarray}\label{eq12}
\frac{\dot{M}_{\rm out}}{\dot{M}_{\rm acc}}
=\frac{4\pi\Omega C\eta \mu m_{p} N_{\rm H}v}{\lbol} \sqrt{\frac{Q_{\rm H}c^3}{4\pi U n_{\rm H}}}.
\end{eqnarray}
The bolometric luminosities for the three quasar SED types are as follows:
$\lbol$ = 4.2 $\lambda_{1500} L_{1500}$ (UV-soft),
$\lbol$ = 6.6 $\lambda_{1500} L_{1500}$ (MF87), 
$\lbol$ = 4.1 $\lambda_{1500} L_{1500}$ (HE 0238).

From Eq.~\ref{eq10}, the kinetic-to-bolometric luminosity ratio can be written as
\begin{eqnarray}\label{eq13}
\frac{\dot{E_{\rm k}}}{\lbol}
=\frac{2\pi\Omega C\mu m_{p} N_{\rm H}v^3}{\lbol} \sqrt{\frac{Q_{\rm H}}{4\pi U n_{\rm H}c}}.
\end{eqnarray}
The $\dot{E_{\rm k}}/\lbol$ distribution can be derived from
the distributions of $CQ_{\rm H}^{1/2}N_{\rm H}v^3\lbol^{-1}$ and $n_{\rm H}$.

We use the skewed Gaussian functions,
\begin{eqnarray}\label{eq14}
N(x)=G(p_1,p_2,p_3)\left\{ 1+{\rm erf}\left[p_{4}(x-p_{1})\right]\right\}
\end{eqnarray}
to model the distribution of $Q_{\rm H}$, $CQ_{\rm H}^{1/2}N_{\rm H}v$, $CQ_{\rm H}^{1/2}N_{\rm H}v\lbol^{-1}$
and $CQ_{\rm H}^{1/2}N_{\rm H}v^3\lbol^{-1}$, where the $p_{1}$ is the mean of Gauss, $p_{2}$ is the standard deviation
of Gauss, $p_{3}$ is the amplitude of Gauss and $p_{4}$ is the coefficient of skewness.

\subsection*{Data availability.}
The data that support the plots within this paper and other findings of this study are available from the corresponding 
author upon reasonable request.

\clearpage

\end{methods}



\noindent \textbf{\huge Supplementary Information}
\addtocounter{figure}{-2} 
\begin{figure*}
\renewcommand{\figurename}{Supplementary Figure}
\centering
\includegraphics[height=13.cm]{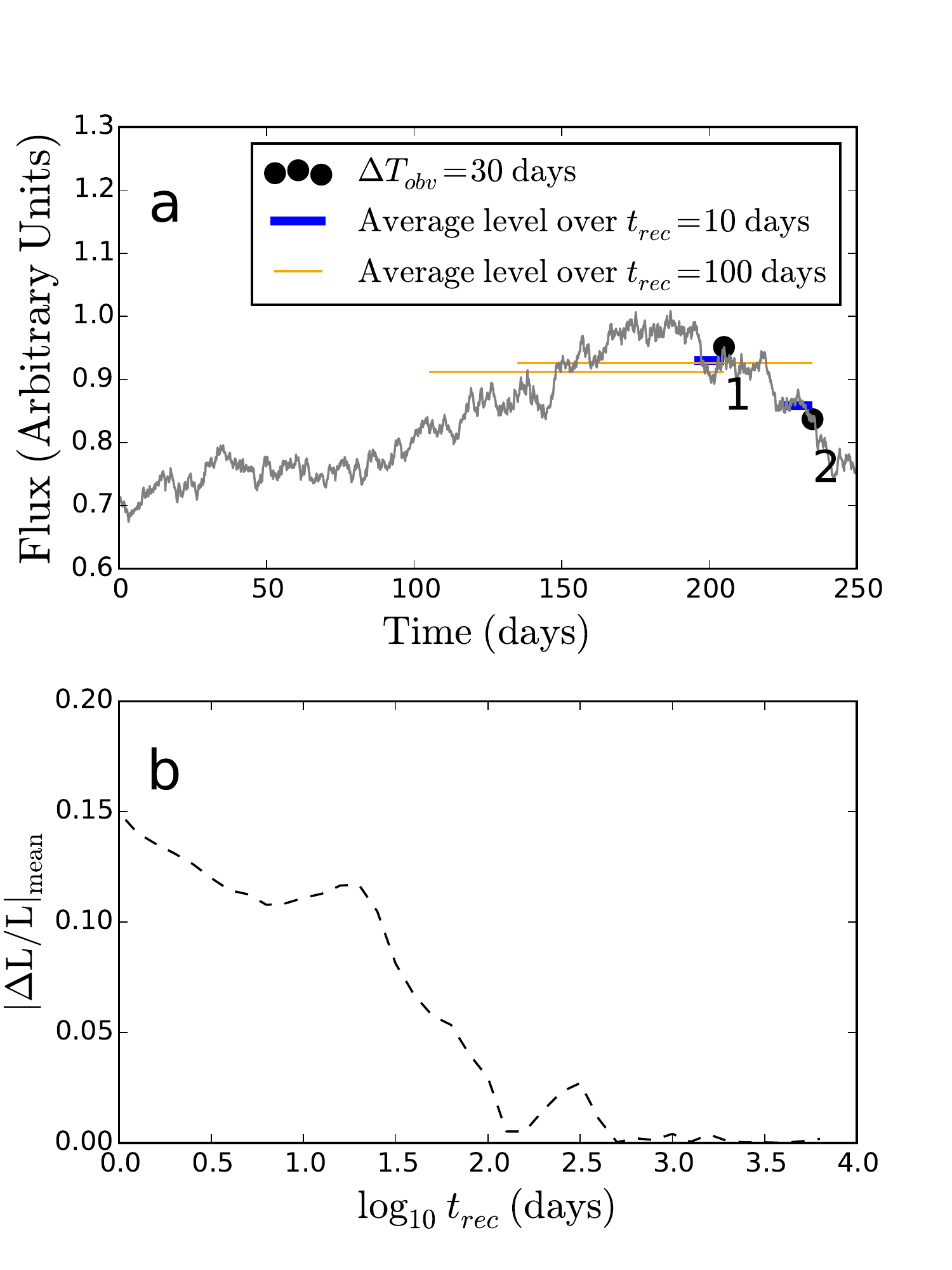}
\caption{\textbf{The change in the average intensity of the ionizing continuum.} 
\textbf{a}, the gray line is an example of the light curve at 1500\AA, generated from the
DRW model. The black points mark the fluxes at two observations with 30 days time interval.
The blue/orange line is the average intensity of ionizing continuum over a recombination
timescale $t_{r}$ 10/100 days before the two observations.
The change in the average intensity of ionizing continuum over the 100 days is obviously smaller than that of 10 days.
\textbf{b}, the change in the average intensity of ionizing continuum between the two observations decreases
when the $t_{r}$ of gas is increased.}
\label{lightcurve}
\end{figure*}

\begin{figure*}
\renewcommand{\figurename}{Supplementary Figure}
\centering
\includegraphics[height=12.cm]{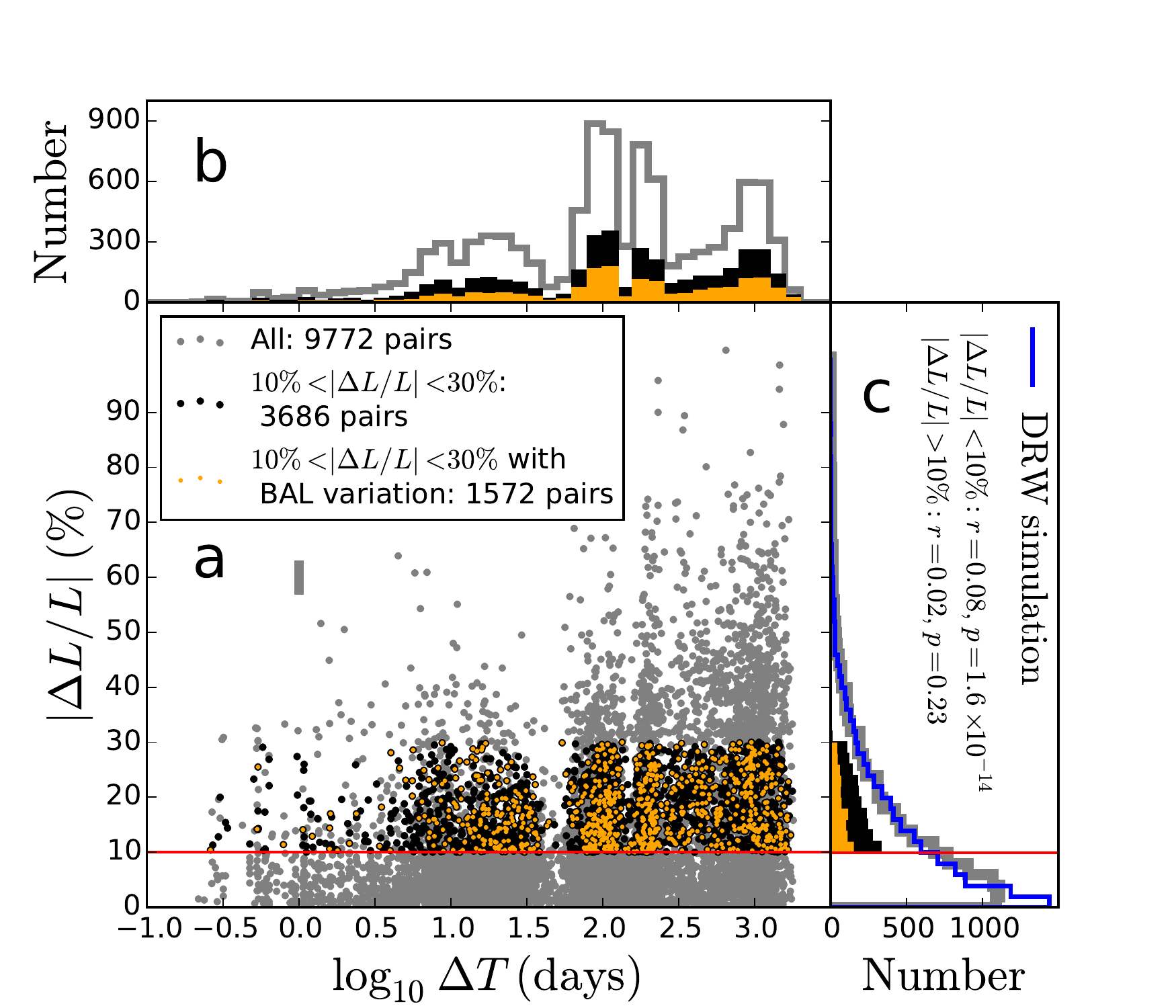}
\caption{\textbf{Properties of the BAL quasar sample.} \textbf{a}, the continuum (1500\AA) variation amplitudes $|\Delta L/L|$ versus the rest time interval $\log_{10}\Delta T$: 9772 spectral pairs of 1728 BAL quasars in the whole sample (gray points);
3686 spectra pairs from 915 quasars with $10\%< |\Delta L/L| <30\%$ (black points); 1572 pairs of spectra from 432 BAL quasars with \civ~BAL variations (orange points). The gray vertical line marks the amplitude of spectrophotometric uncertainties of 6\% level\cite{shen2014}.
The red horizontal line marks the level of $|\Delta L/L|=10\%$.
\textbf{b}, the distributions of the rest time interval $\log_{10}\Delta T$. The colors are
the same with \textbf{a}. \textbf{c}, the distributions of the amplitudes of continuum variations $|\Delta L/L|$ for the whole sample (gray line) and the DRW model (blue line) of a typical quasar ($\mbh=10^9\msun$, $M_{i}=-26$) are consistent at $|\Delta L/L| \ge$ 10\% 
(Kolmogorov-Smirnov test: $r=0.02,~p=0.23$).
These two distributions with $|\Delta L/L|<$ 10\% are significantly different ($r=0.08,~p=1.6\times 10^{-14}$).}
\label{fig:obvtime}
\end{figure*}

\addtocounter{figure}{-1} 
\begin{figure*}
\renewcommand{\figurename}{Supplementary Figure}
\centering
\includegraphics[height=10.cm]{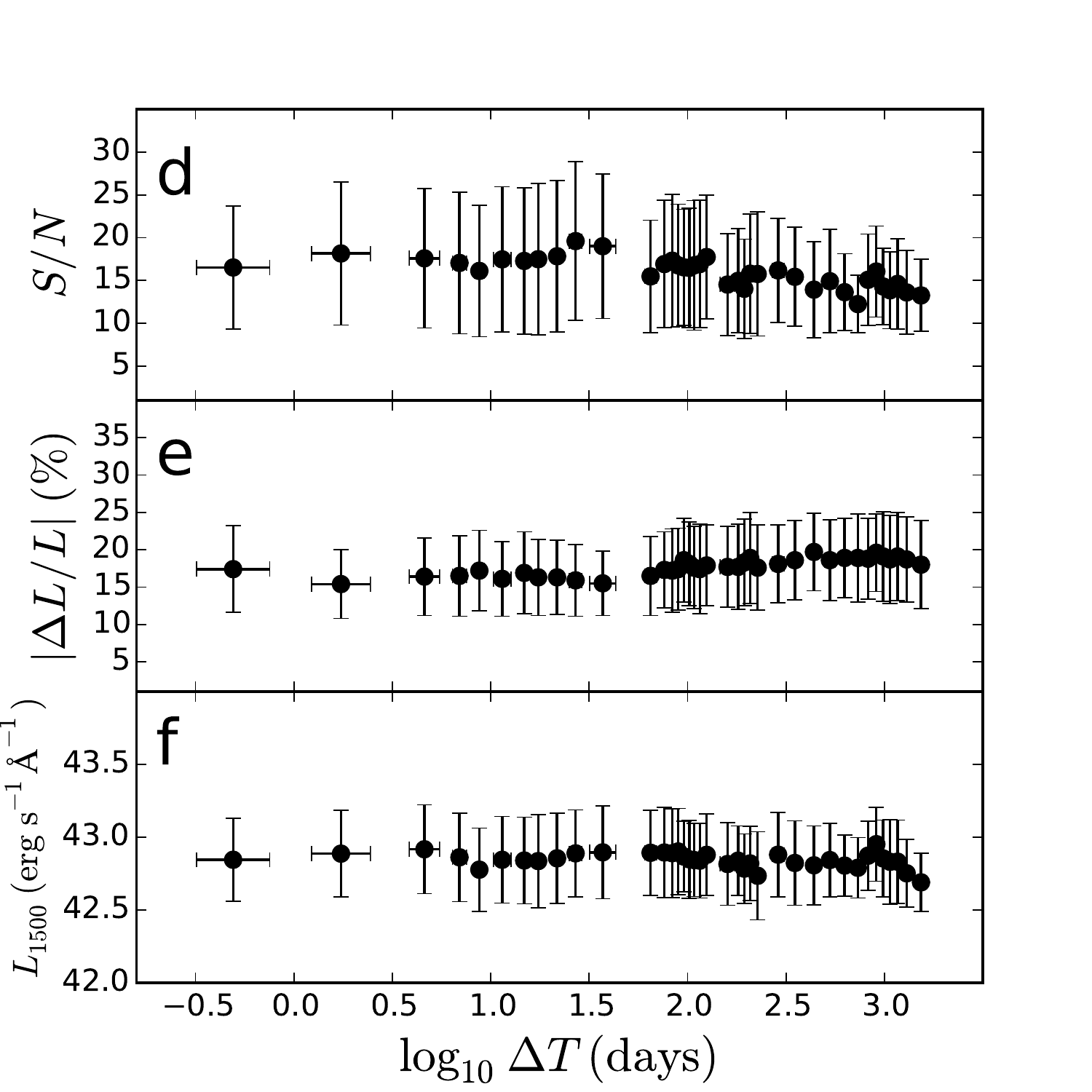}
\caption{\textbf{Properties of the BAL quasar sample.}
\textbf{d}, the mean and standard deviation of the S/N of SDSS $g$ band of the spectra for 3686 pairs from the 915 
quasars at different time bins. \textbf{e}, the mean and standard deviation of the amplitudes of continuum variations 
$|\Delta L/L|$ at 1500\AA. \textbf{f}, the mean and standard deviation of the monochromatic luminosities at 1500\AA. 
All these quantities are nearly constants in all time intervals.}
\label{fig:obvtime}
\end{figure*}

\begin{figure*}
\renewcommand{\figurename}{Supplementary Figure}
\centering
\includegraphics[height=9.cm]{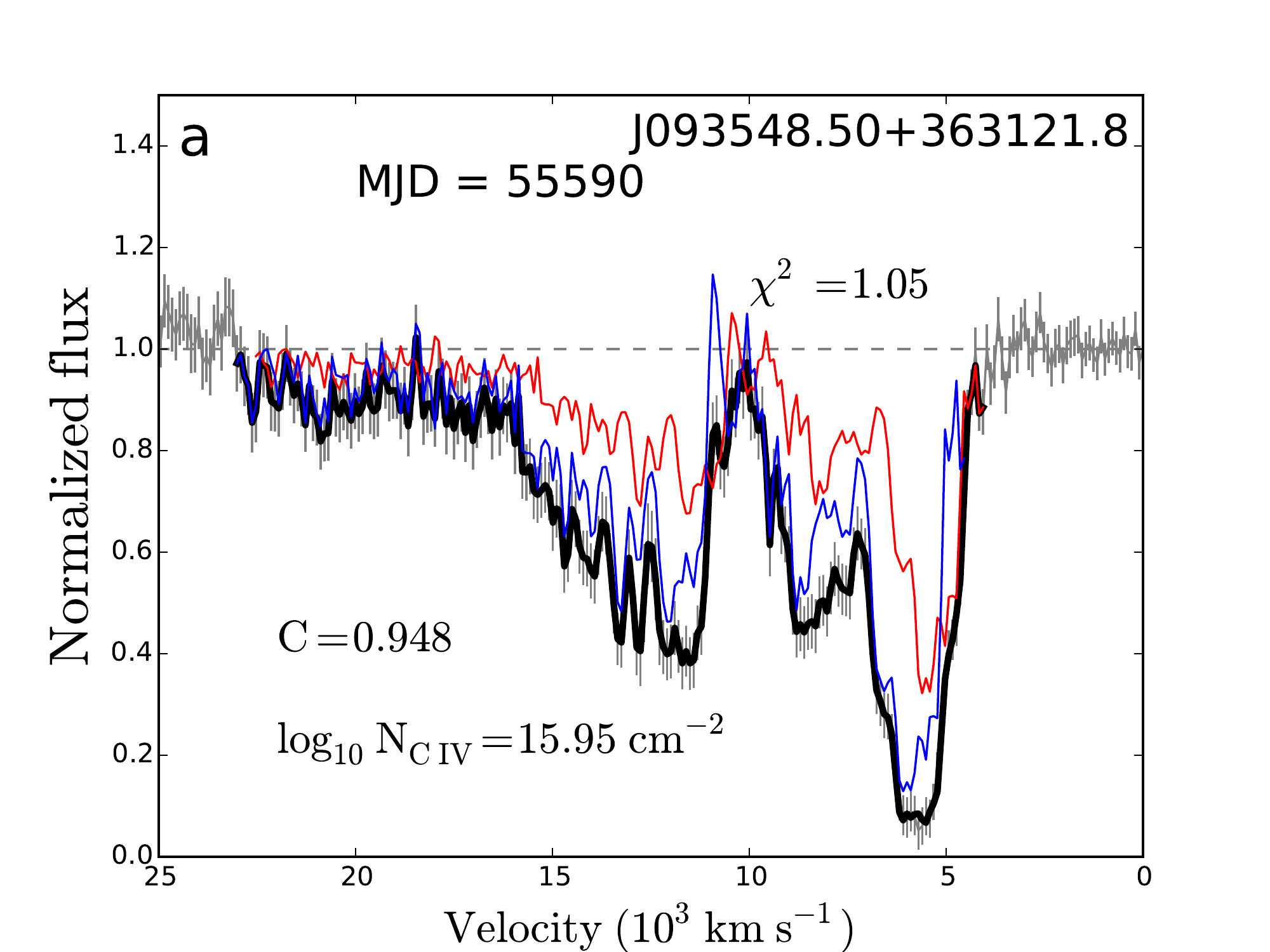}
\includegraphics[height=9.cm]{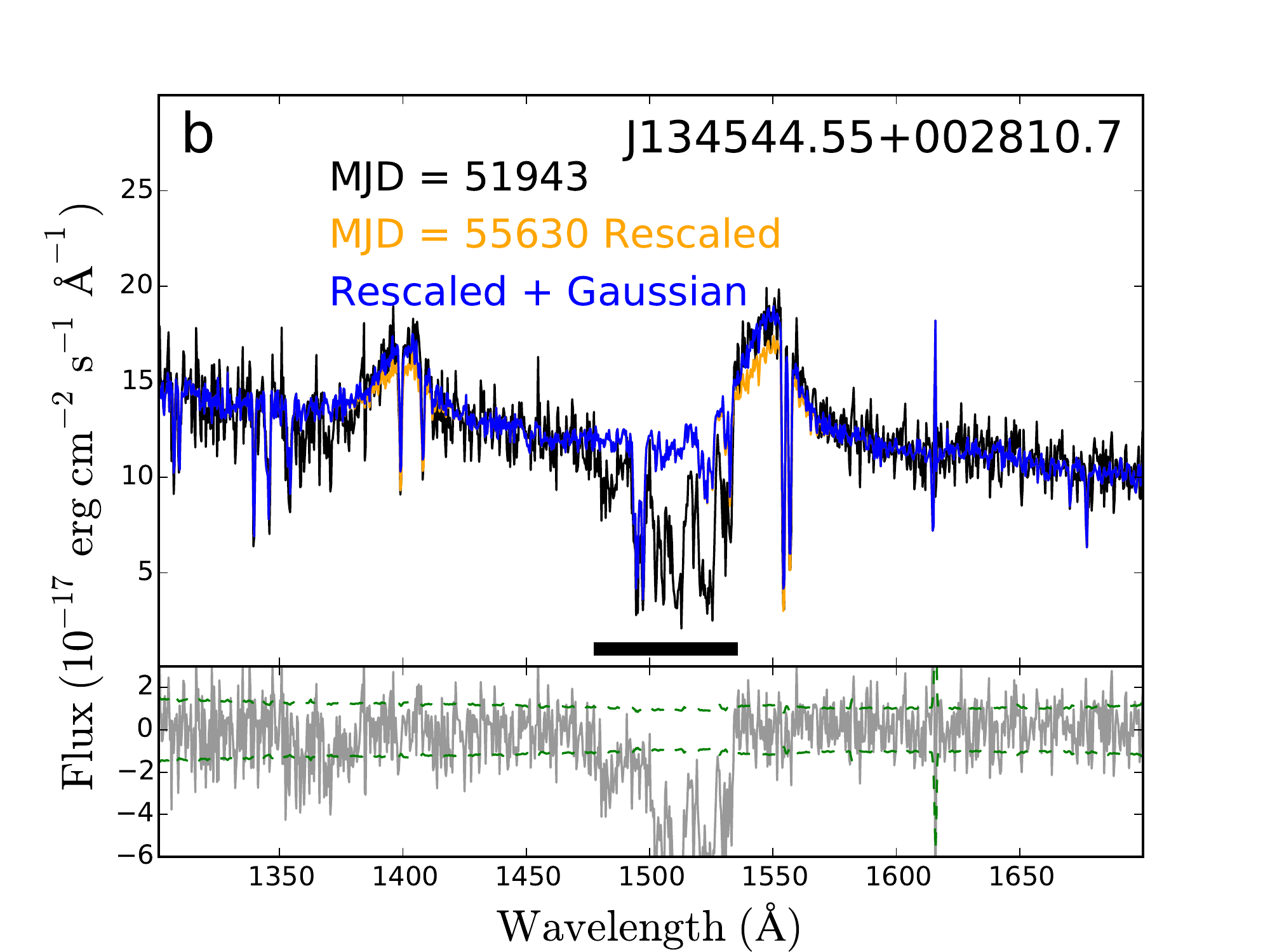}
\caption{\textbf{Fit of the \civ~BAL troughs and identification of variable absorption Lines.}
\textbf{a}, an example of the \civ1548.2,1550.8\AA~doublet blended BAL trough.
The gray line is the observational spectrum normalized by the continuum. 
The horizontal dashed line marks the level of the continuum.
The red/blue line is the best fitted red/blue component of the doublets, and the thick black one is the
product of the doublets. 
\textbf{b}, an example of matching the reference spectrum to another spectrum (in black) 
by multiplying the reference spectrum with a double power-law described in the text (the orange line). 
The blue curve represents the one with additional Gaussians to account for the change of
the emission line equivalent width. The black horizontal line represents the varied
region of \civ~BAL. The residuals of fits (solid line) and the combined
spectrum uncertainties (dashed line) are plotted in the lower panel.  }
\label{variregion}
\end{figure*}

\begin{figure*}
\renewcommand{\figurename}{Supplementary Figure}
\centering
\includegraphics[height=12.cm]{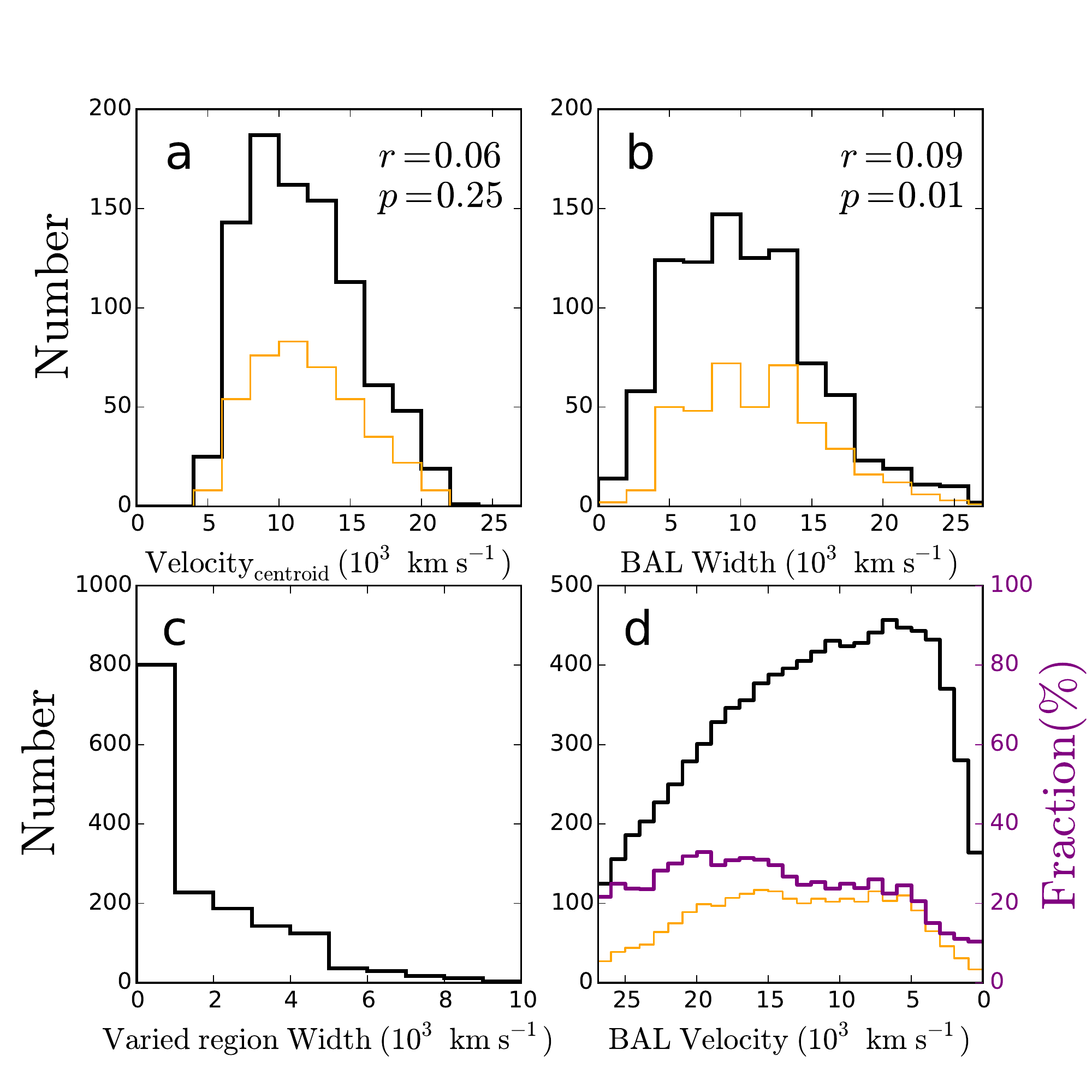}
\caption{\textbf{Distributions of the parameters of the \civ~BAL troughs.} The black line is the 915
quasar with $10\%< |\Delta L/L| <30\%$ and the orange one is for the 432 quasar with varied BALs.
\textbf{a}, the distributions of the weighted centroid velocities. There is no significant difference between
the black and orange (Kolmogorov-Smirnov test: $r=0.06,~p=0.25$).
\textbf{b}, the distributions of BAL trough widths. There is a weak difference between
the black and orange ($r=0.09,~p=0.01$).
\textbf{c}, the distributions of varied BAL region widths for 1572 spectral pairs.
\textbf{d}, the number of occurrences of BAL absorption (black) and variable region (orange) at different velocity bin.
The purple one is the percentage of BAL variability, i.e., the ratio of the orange one to the black one. }
\label{fig:trough}
\end{figure*}

\begin{figure*}
\renewcommand{\figurename}{Supplementary Figure}
\centering
\includegraphics[height=9.cm]{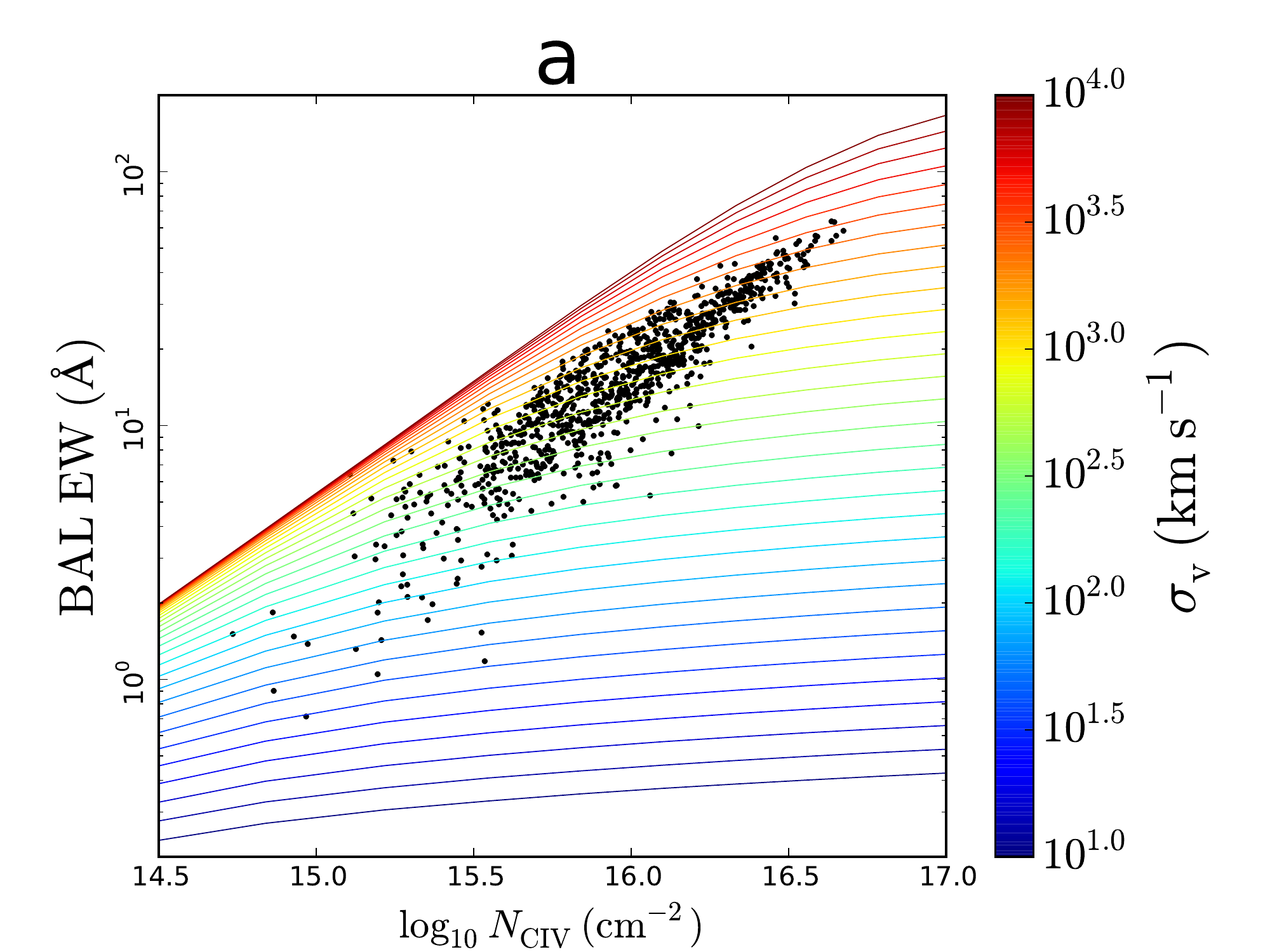}
\includegraphics[height=9.cm]{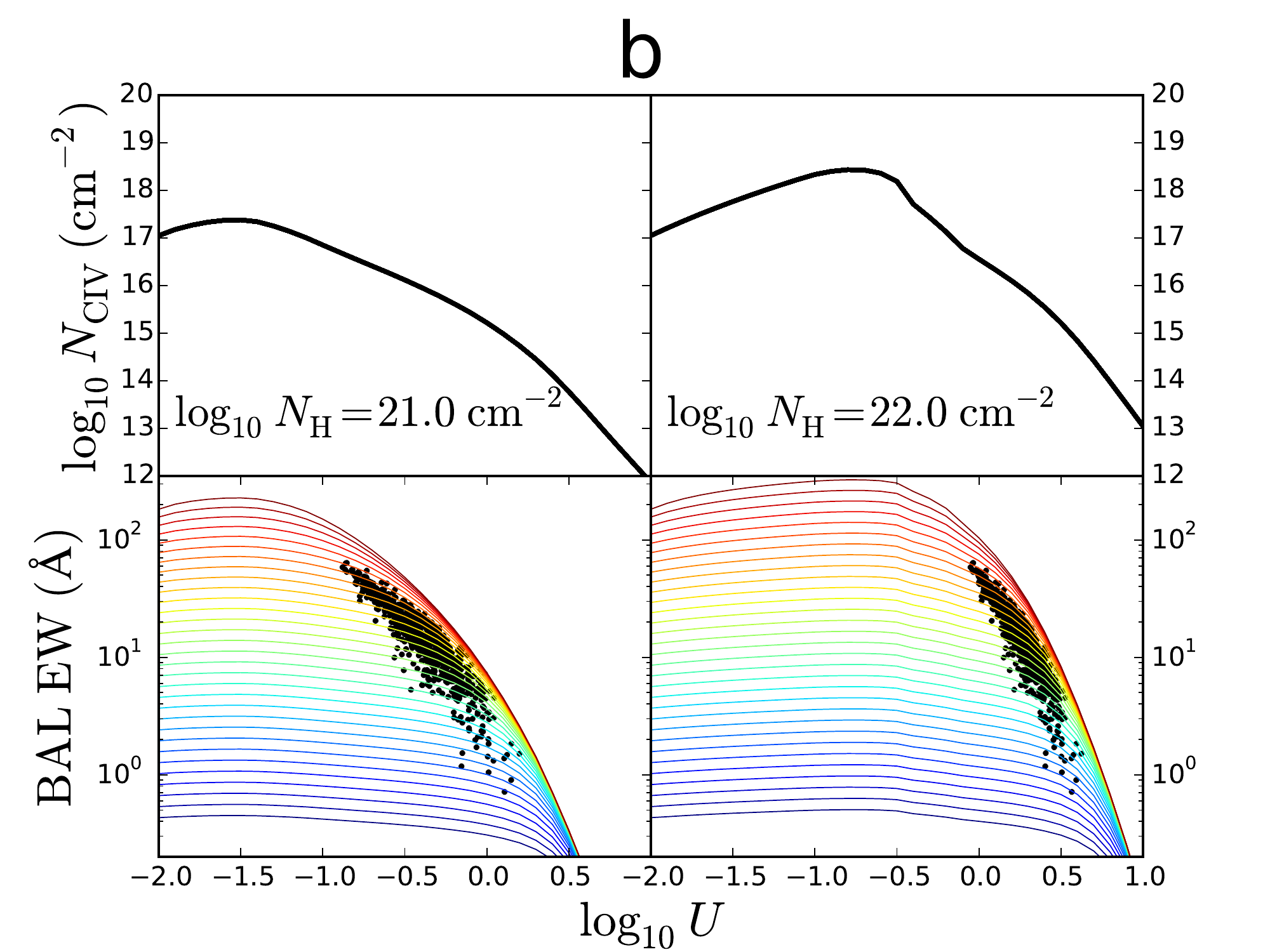}
\caption{\textbf{Simulation of the BAL EW response to the variation of ionizing flux.} 
\textbf{a}, the black dots are the measured \civ~BAL EW and column density $N_{\rm CIV}$ for the 915 quasars.
The typical $1\sigma$ error bar of $\log_{10}\rm BAL\ EW$ is 0.014 $\rm dex$.
The typical $1\sigma$ error bar of $\log_{10} N_{\rm CIV}$ is 0.03 $\rm dex$.
The color lines are the BAL EW curves as a function of the $N_{\rm CIV}$ at different widths $\rm \sigma_{v}$.
\textbf{b}, the response of $N_{\rm CIV}$ and BAL EW to the variations of ionization parameters $\log_{10} U$ 
using Cloudy~\cite{ferland2013} at the hydrogen column density 
$\log_{10}N_{\rm H}$ = 21 $\rm cm^{-2}$ and 22 $\rm cm^{-22}$. The BAL EW will respond to the $\log_{10} U$ along 
each color line at different widths $\rm \sigma_{v}$.}
\label{Velocity}
\end{figure*}

\begin{figure*}
\renewcommand{\figurename}{Supplementary Figure}
\centering
\includegraphics[height=11.2cm]{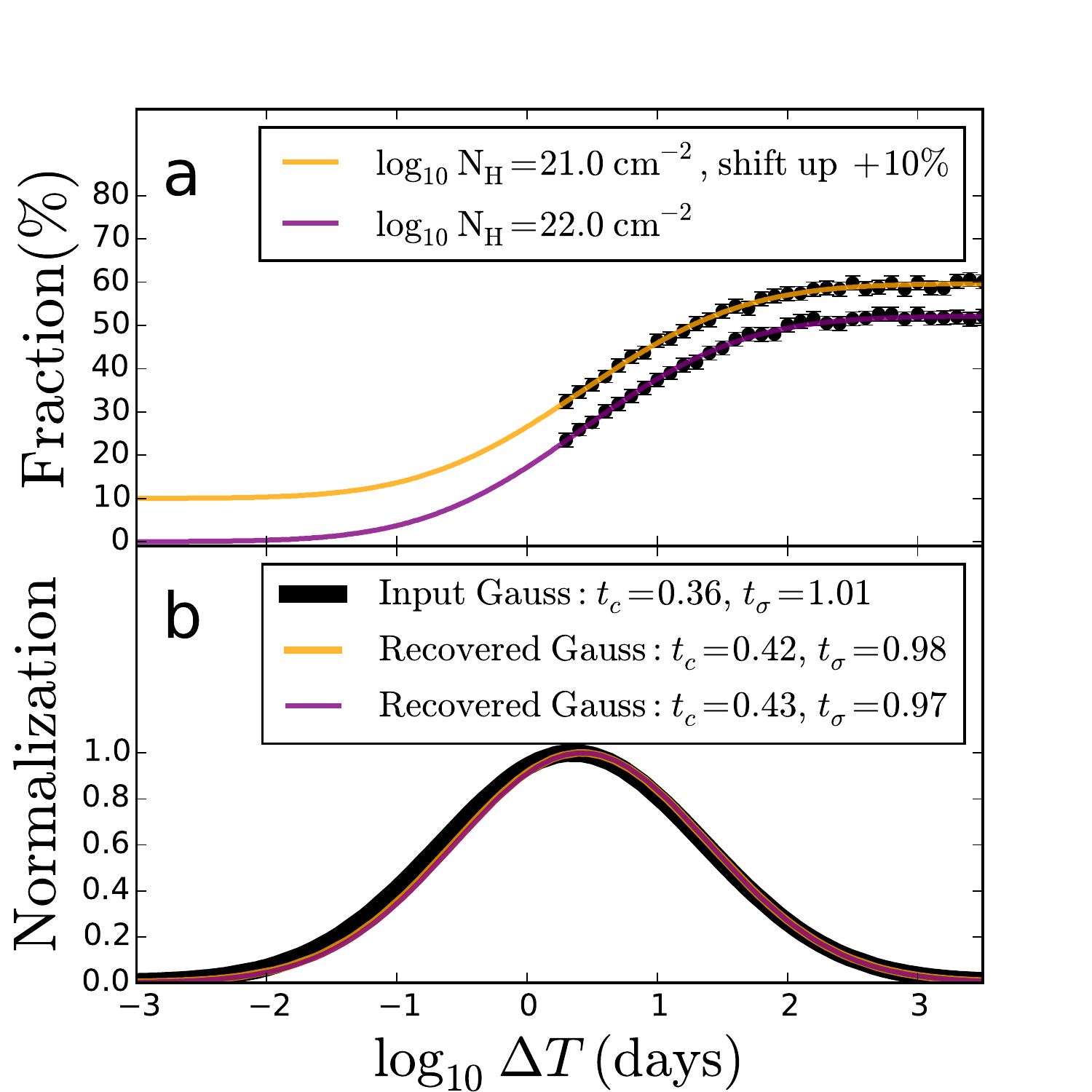}
\caption{\textbf{Simulation result.} \textbf{a}, the black points with error bars are the $F(\log_{10}\Delta T)$ with
$1\sigma$ uncertainty measured from the mock sample. 
We use the CDF of a Gaussian distribution (see Eq.~2) to model the fraction curves. 
The orange and purple lines are the best fitted models for $\log_{10}N_{\rm H}$ = 21 $\rm cm^{-2}$ and 
$\log_{10}N_{\rm H}$ = 22 $\rm cm^{-2}$, respectively.
\textbf{b}, the input (black thick line) and recovered (orange for  $\log_{10}N_{\rm H}$ = 21 $\rm cm^{-2}$ and purple
for $\log_{10}N_{\rm H}$ = 22 $\rm cm^{-2}$) Gausses. The mean and standard deviation of the
input Gaussian distribution are $t_{c}=0.36\pm0.14$ and $t_{\sigma}=1.01\pm0.22$.
The best fit parameters of the recovered Gauss for $\log_{10}N_{\rm H}$ = 21 $\rm cm^{-2}$
are: $t_{c}=0.42\pm0.03$ and $t_{\sigma}=0.98\pm0.06$.
For $\log_{10}N_{\rm H}$ = 22 $\rm cm^{-2}$, the best fit parameters
are: $t_{c}=0.43\pm0.03$ and $t_{\sigma}=0.97\pm0.06$.
The parameters of the recovered Gausses are consistent with the parameters of the input Gauss within 1$\sigma$ uncertainty.}
\label{fractioncurve}
\end{figure*}

\begin{figure*}
\renewcommand{\figurename}{Supplementary Figure}
\centering
\includegraphics[height=8.5cm]{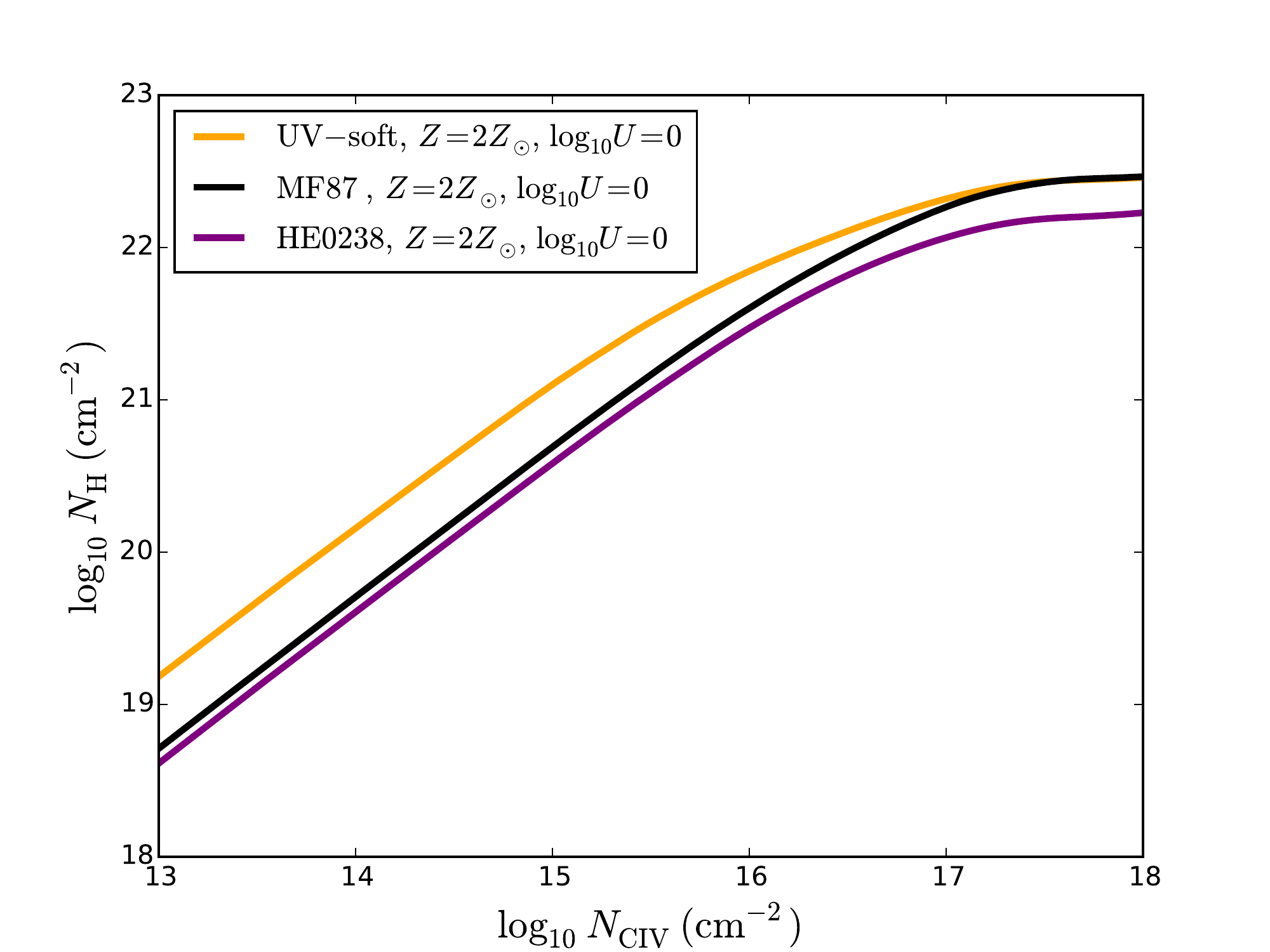}
\caption{
\textbf{The hydrogen column density $N_{\rm H}$ at the ionization parameter ${\log_{10}}~U = 0$.} 
The hydrogen column density $N_{\rm H}$ of all the outflows can be estimated from 
the \civ~column density $N_{\rm CIV}$. The orange, black and purple lines represent the UV-Soft,
MF87 and HE0238 SEDs, respectively.}
\label{fig:civnh}
\end{figure*}

\begin{figure*}
\renewcommand{\figurename}{Supplementary Figure}
\centering
\includegraphics[height=13.5cm]{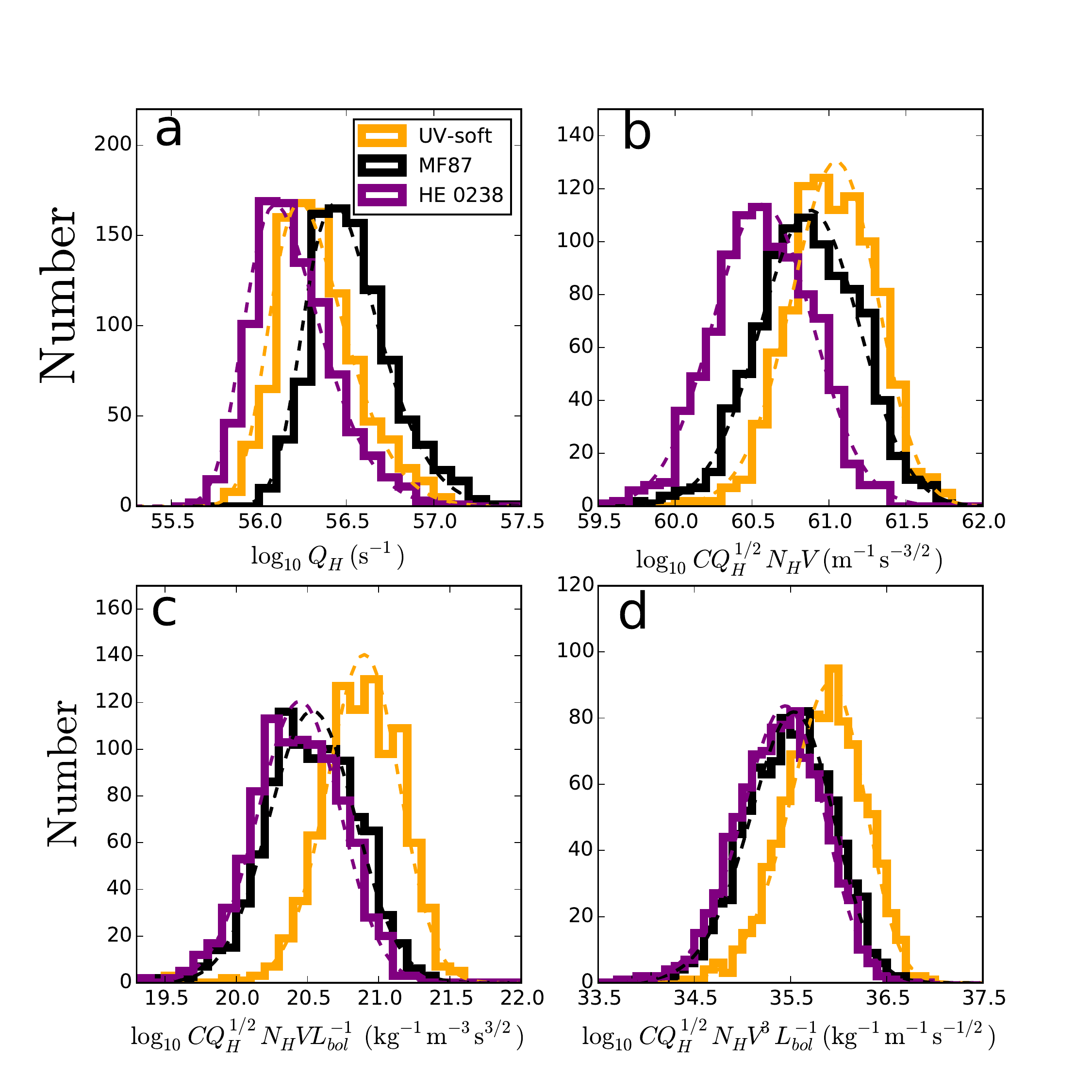}
\caption{\textbf{Supplementary distributions at ${\log_{10}}~U = 0$.} We use the skewed Gaussian functions
to model the distributions: \textbf{a}, $Q_{\rm H}$; \textbf{b}, $CQ_{\rm H}^{1/2}N_{\rm H}v$; 
\textbf{c}, $CQ_{\rm H}^{1/2}N_{\rm H}v\lbol^{-1}$; 
\textbf{d}, $CQ_{\rm H}^{1/2}N_{\rm H}v^3\lbol^{-1}$. The dashed lines are the best fit results.
The orange, black and purple lines represent the UV-Soft,
MF87 and HE0238 SEDs, respectively.}
\label{fig:final}
\end{figure*}

\end{document}